\documentclass[prb,twocolumn]{revtex4-1}
\usepackage{amssymb,amsfonts,amsmath}
\usepackage{adjustbox}
\usepackage{hyperref}

\begin{document}
\title{Topology of atomically thin soft ferroelectric membranes at finite temperature}
\author{John W. Villanova}
\email{jvillano@uark.edu}
\affiliation{Department of Physics, University of Arkansas, Fayetteville, Arkansas 72701, United States}
\author{Pradeep Kumar}
\affiliation{Department of Physics, University of Arkansas, Fayetteville, Arkansas 72701, United States}
\author{Salvador Barraza-Lopez}
\email{sbarraza@uark.edu}
\affiliation{Department of Physics, University of Arkansas, Fayetteville, Arkansas 72701, United States}

\begin{abstract}
One account of two-dimensional (2D) structural transformations in 2D ferroelectrics predicts an evolution from a structure with Pnm2$_1$ symmetry into a structure with
square P4/nmm symmetry and is consistent with experimental evidence, while another argues for a transformation into a structure with rectangular Pnmm symmetry. An analysis
of the assumptions made in these models is provided here, and six fundamental results concerning these transformations are contributed as follows: (i) Softened phonon modes
produce rotational modes in these materials. (ii) The transformation to a structure with P4/nmm symmetry occurs at the lowest critical temperature $T_c$. (iii) The
hypothesis that one unidirectional optical vibrational mode underpins the 2D transformation is unwarranted. (iv) Being successively more constrained, a succession of
critical temperatures ($T_c<T_c'<T_c''$) occurs in going from molecular dynamics calculations with the NPT and NVT ensembles onto the model with unidirectional
oscillations. (v) The choice of exchange-correlation functional impacts the estimate of the critical temperature. (vi) Crucially, the correct physical picture of these
transformations is one in which rotational modes confer a topological character to the 2D transformation via the proliferation of vortices.
\end{abstract}

\maketitle

\date{today}

\section{Introduction}

Group-IV monochalcogenide monolayers (MLs) \cite{Kai,KaiPRL} are two-atom thick ferroelectric semiconducting membranes \cite{fei_apl_2015_ges_gese_sns_snse} which command
interest for nonlinear optical applications \cite{wang_nanolett_2017_gese,b4} and all-electric non-volatile memories \cite{kai3} originating from their lack of inversion
symmetry and their in-plane polarization. A task for theory is to develop predictive quantitative estimates of the critical temperature ($T_c$) at which a ferroelectric to
paraelectric transformation takes place, and determining $T_c$ in the simple case of freestanding samples remains an unsettled business. The purpose of the present work is to
distinguish results from within two models, in order to establish the proper framework to understand their structural properties at finite temperature ($T$). The study is
explicitly carried out for SnSe MLs and it applies to group-IV monochalcogenide MLs with a rectangular ground state unit cell (u.c.).

\begin{figure}[t]
\begin{center}
\includegraphics[width=0.46\textwidth]{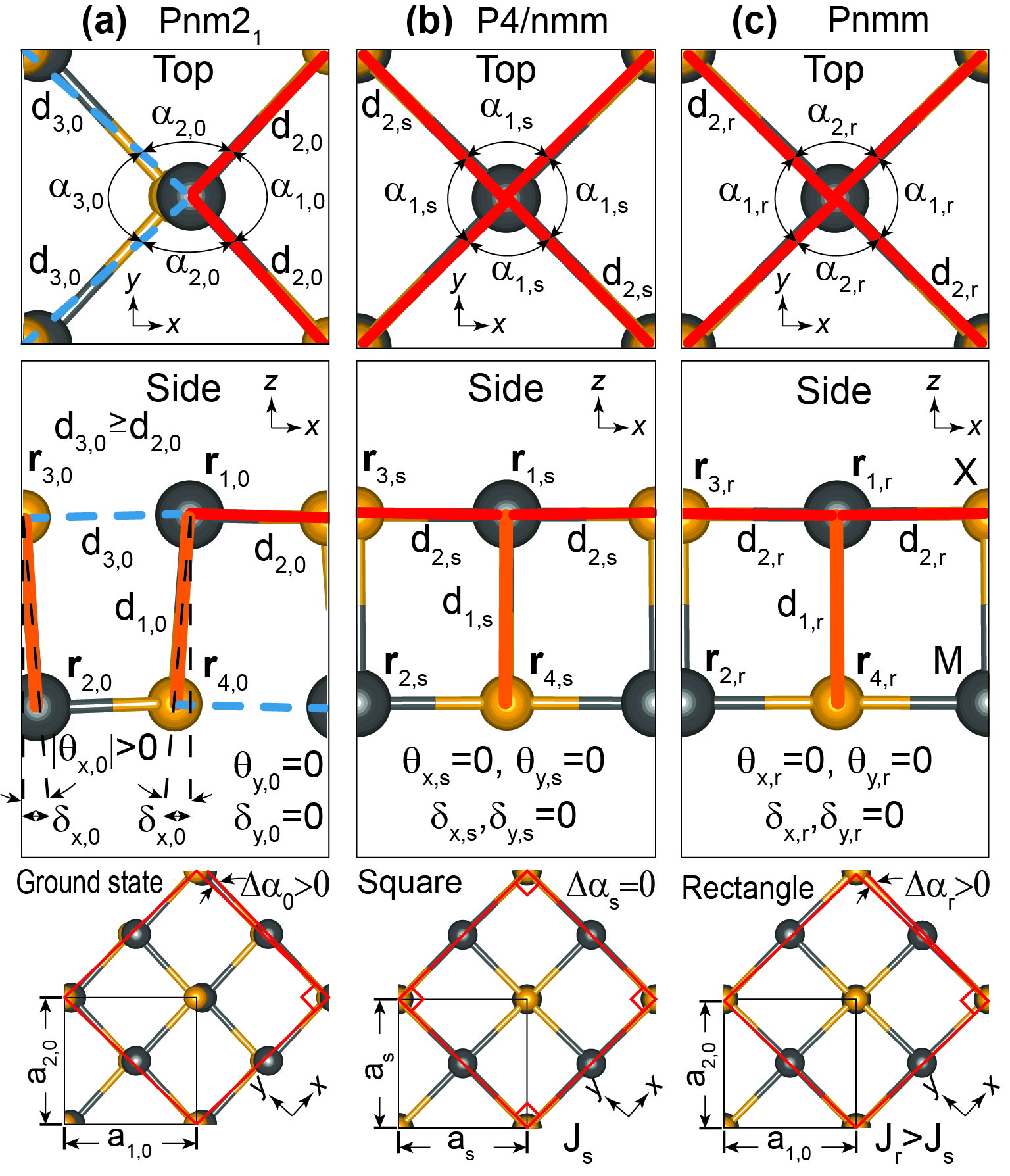}
\caption{(a) Ferroelectric rectangular ground state u.c.~(with Pnm$2_1$ group symmetry) of a SnSe ML. The Sn (Se) atom is gray (orange); $\Delta \alpha_0 > 0$. (b)
Paraelectric square (P4/nmm) phase in which $a_1=a_2=a_s$, $d_{3,s}=d_{2,s}$, $\alpha_1=\alpha_2=\alpha_3=\alpha_{1,s}$, $\delta_{x,s}=\delta_{y,s}=0$, and $\Delta
\alpha_s=0$. (c) Paraelectric rectangular (Pnmm) phase in which $a_1$ and $a_2$ retain ground state magnitudes (so $\Delta \alpha_r = \Delta \alpha_0$), $d_{3,r}=d_{2,r}$,
$\alpha_1=\alpha_3=\alpha_{1,r}\ne \alpha_{2,r}$, $\delta_{x,r}=\delta_{y,r}=0$. The energy ($J_s$ or $J_r$) in turning from structure (a) into (b) or (c) is indicated, too.
(Reproduced from Ref.~\cite{shiva}.)}\label{fig:figure1}
\end{center}
\end{figure}

The original procedure to estimate $T_c$ in these transformations \cite{Mehboudi2016} contained three ingredients. (i) The four degenerate ferroelectric structural
ground states that can be generated out of a rectangular (Pnm2$_1$ \cite{rodin_prb_2016_sns}) u.c.~once the two in-plane principal axes ($x$ and $y$ in Fig.~\ref{fig:figure1})
are established: looking at Fig.~\ref{fig:figure1}(a), the fourfold degeneracy means that the total energy of the SnSe ML remains unchanged by a reflection with respect to the
$y-$axis and/or by a swap of $x-$ and $y-$coordinates \cite{Mehboudi2016, otherarticle, other3}. (ii) The second ingredient is an energy barrier $J_s$ separating these ground
states. This energy barrier is obtained as the energy difference per u.c.~between a paraelectric atomistic structure with enhanced symmetry and the fourfold degenerate ground
state u.c.~shown in Fig.~\ref{fig:figure1}(a); the (tetragonal) paraelectric u.c.~with fourfold rotational symmetry (P4/nmm) \cite{Mehboudi2016,otherarticle,other3} is shown in
Fig.~\ref{fig:figure1}(b). (iii) The third ingredient is an order-by-disorder Potts-model \cite{potts} that was used to create the finite-temperature behavior of these MLs.
Potts models are well-known tools within soft matter and statistical physics to deal with structural transformations in two dimensions. A simple analytical relation exists
between the energy barrier $J_s$ and $T_c$ in the Potts model \cite{potts}: $T_c=\frac{2J_s}{1.76 k_B}$, where $k_B$ is the Boltzmann constant. This applies when the number
of degenerate ground states (expressed as $r$ in the original article) is equal to four as in the case at hand. Experimental work on SnTe MLs on a graphitic substrate showed
evidence for a structural transformation at finite temperature with an order parameter known as the {\em rhombic distortion angle} $\Delta\alpha$ turning zero at $T_c$
\cite{Kai}.

Subsequent theoretical work \cite{fei_prl_2016} settles on an orthorhombic paraelectric u.c.~and it also differs in its underlying assumptions for fundamental reasons that will be explicitly identified. For instance, it has been shown that $\Delta\alpha=0$ holds only if $a_1=a_2$ \cite{other4,Mehboudi2016}. Additionally, a recent analysis shows that the paraelectric u.c.s~in Figs.~\ref{fig:figure1}(b) and \ref{fig:figure1}(c) are completely determined using different numbers of structural constraints \cite{shiva}: the structure obeying the experimental order parameter $\Delta\alpha=0$ at $T_c$ [Fig.~\ref{fig:figure1}(b)] has fewer structural constraints than the structure championed by Ref.~\cite{fei_prl_2016} and seen in Fig.~\ref{fig:figure1}(c), and should in principle be reached at a lower critical temperature accordingly.

This article was written in a gradual and thorough manner in an attempt for clarity of exposition. After introducing the numerical methods in Sec.~\ref{sec:methods}, the
assumptions made by the model in Ref.~\cite{fei_prl_2016}--henceforth labeled unidirectional optical vibration (UOV) model--are provided in Sec.~\ref{sec:iii.a}. The
structural order parameters used to understand the different finite-$T$ phases of group-IV monochalcogenide MLs are indicated in Sec.~\ref{sec:iii.b}, and the number
of free parameters necessary to reach three different structural phases are indicated in Sec.~\ref{sec:iii.c}. The main points of Secs.~\ref{sec:iii.b} and \ref{sec:iii.c}
can be found in Refs.~\cite{other4} and \cite{shiva}; they were included for a self-contained discussion since they have a bearing on the hypotheses of the UOV
model.

Section \ref{sec:iii.d} contains the eigenvalue and eigenvector spectrum for the ground state unit cell (symmetry group Pnm2$_1$), and the unit cells with P4/nmm and Pnmm
symmetry, to which different finite-temperature transformations lead. The comparative finite-$T$ evolution of a SnSe monolayer as obtained within the (isobaric-isothermal) NPT and the (canonical) NVT ensembles and with the UOV model is presented in Sec.~\ref{sec:iii.e}; a discussion of limits in which the NVT and NPT ensembles agree is provided at this point, too. Along the way, a case for the existence of rotational modes in these materials will be made, and it will be shown that the transformation to a structure with P4/nmm symmetry occurs at the lowest critical temperature $T_c$, making it the most physically viable. A succession of critical temperatures ($T_c<T_c'<T_c''$) in going from molecular dynamics (MD) calculations with the NPT and NVT ensembles onto the UOV model will be obtained. The effect of exchange-correlation (XC) functional on interatomic forces alters predictions of $T_c$, too. MD data are shown to violate one hypothesis of the UOV model (angle covariant approximation) in Sec.~\ref{sec:iii.f} and the unidirectional vibration hypothesis in Sec.~\ref{sec:iii.g}.

This article owes its title to the fact that the physics of the structural transformation features topological vortices due to the time-evolving connectivity of the lattice,
and it builds its case up to these results, which are presented in Sec.~\ref{sec:iii.h}.

Conclusions are presented in Section \ref{sec:conclusions}. A detailed description of the process to parameterize the UOV model can be found in Appendix
\ref{sec:appendixA}, and a thorough exposition of relevant group symmetries is provided as Appendix \ref{sec:appendixB}.

\section{Methods}\label{sec:methods}
We report density functional theory calculations on SnSe MLs employing the \emph{SIESTA} code \cite{siesta} with vdW-DF-cx van der Waals corrections \cite{soler,BH}.
Additional calculations were performed with the PBE XC functional \cite{PBE} for direct comparison with Ref.~\cite{fei_prl_2016}. The evolution of order parameters is studied
through \emph{ab initio} MD on $16 \times 16$ supercells containing 1024 atoms for over 20,000 fs (with 1.5 fs resolution), employing the isothermal-isobaric (NPT) and canonical (NVT) ensembles which lead to structural transformations into the P4/nmm or Pnmm phases, respectively. A parametrization of the model in Ref.~\cite{fei_prl_2016} was also created for comparison.

\section{Results and discussion}

\subsection{Hypotheses of the UOV model}\label{sec:iii.a}
The UOV model \cite{fei_prl_2016} for 2D structural transformations relies on five assumptions, and a choice of XC functional:
\begin{enumerate}
\item{}({\bf UOV 1}): There are two (instead of four \cite{Mehboudi2016}) structural ground state u.c.s.
\item{}({\bf UOV 2}): The paraelectric u.c.~is rectangular. Details concerning the appropriate symmetry group are provided in Appendix \ref{sec:appendixB}.
\item{}({\bf UOV 3}): The structural transformation is driven by a single optical vibrational mode, that oscillates unidirectionally. (These are {\em two} consecutive assumptions.)
\item{}({\bf UOV 4}): The {\em angle-covariant approximation} forces the two electric dipole moments per unit cell to always have the same angle ($\theta_x$) of deviation
    from the out-of-plane direction.
\item{}({\bf UOV 5}): The finite-$T$ dynamics of neighboring unit cells were parameterized in a mean-field approximation.
\item{}({\bf UOV 6}): Parameters were fit against monolayer structures obtained with the PBE XC functional \cite{PBE}.
\end{enumerate}
These six hypotheses will be carefully scrutinized (and some of them discarded) in order to properly understand the differences between MD results and those generated from the UOV model.

\subsection{Order parameters in the ground state and paraelectric structures}\label{sec:iii.b}
Structural order parameters \cite{other4} will be utilized to understand the structural transformation and are discussed here for completeness. The ground state
crystal structure of a SnSe ML (denoted with a subscript ``0'') shown in Fig.~\ref{fig:figure1}(a) has a Pnm2$_1$ symmetry \cite{rodin_prb_2016_sns}. The atoms in this
u.c.~are labeled $\mathbf{r}_{i,0}$ ($i=1,2,3,4$), and a few geometrical order parameters are listed. From top to bottom, they are angles $\alpha_{i,0}$ ($i=1,2,3$)
formed among the upper layer of atoms and distances $d_{1,0}$ (orange solid lines), $d_{2,0}$ (red solid lines), and $d_{3,0}$ (blue dashed lines) between first, second, and
third nearest neighbors, respectively. Additional order parameters include the projection $\delta_{x,0}$ of the orange line onto the horizontal ($x$) direction and the angle
$\theta_{x,0}=\arcsin(\delta_{x,0}/d_{1,0})$ \cite{fei_prl_2016}, which can be seen {\em twice} in Fig.~\ref{fig:figure1}(a), side view. The bottom subplot depicts lattice
parameters $a_{1,0}$, $a_{2,0}$ and the rhombic distortion angle $\Delta\alpha_0$ \cite{Kai,other4}.

\begin{figure*}[tb]
\begin{center}
\includegraphics[width=0.96\textwidth]{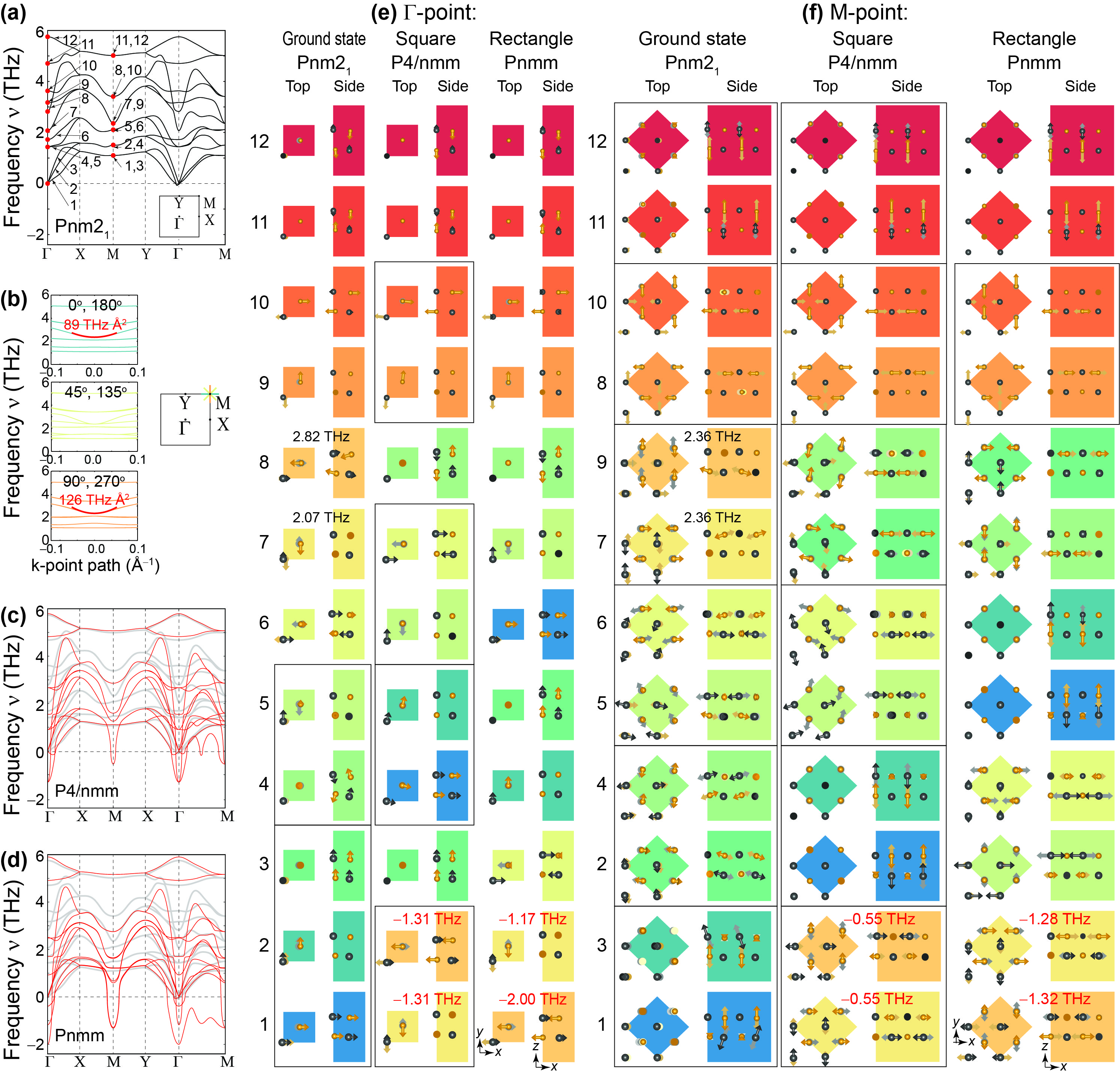}
\caption{(a) Phonon dispersion for the SnSe ML at its ground state (Pnm2$_1$ symmetry). Eigenfrequencies $\nu$ are listed in ascending order, according to their magnitudes
around the $\Gamma-$point. (b) Rotational signatures of degenerate modes 7 and 9 at the $M-$point for all momentum directions are emphasized by red quadratic fits.  (c,d)
Phonon dispersion in the P4/nmm and Pnmm u.c., respectively~[gray curves are those in subplot (a); they help identifying softened phonons]. (e,f) Identification of softened
phonons in P4/nmm and Pnmm phases at the $\Gamma-$ or $M-$point, respectively. Note that soft (optical) {\em mode 7 is not even unidirectional at the $M-$point}, in contradiction to assumption {\bf UOV 3}. Bounding rectangles identify degenerate eigenfrequencies.} \label{fig:figure2}
\end{center}
\end{figure*}

We reported a theoretically determined structural transformation at finite $T$ whereby the u.c.~turns into the {\em tetragonal} P4/nmm structure displayed in
Fig.~\ref{fig:figure1}(b). Such fourfold symmetry (and its tetragonal unit cell) implies that $a_1=a_2=a_s$, $\delta_x=\delta_y=0$, $\Delta\alpha_s=0$, and all angles
$\alpha_{i,s}=90^{\circ}$ \cite{Mehboudi2016,other2}.

Ref.~\cite{fei_prl_2016} reproduced the decay of the intrinsic dipole moment documented in Ref.~\cite{other2} but considered a 2D transformation into the {\em
orthorhombic}, rectangular paraelectric structure with Pnmm symmetry seen in Fig.~\ref{fig:figure1}(c) in which angles $\alpha_i$ ($i=1,2,3$) take on the two values
$\alpha_{1,r}$ and $\alpha_{2,r}$, and $\Delta\alpha_r \ne 0$. The assumption that the paraelectric structure is rectangular is hypothesis {\bf UOV 2} above.

The symmetries of these theoretical predictions are not commensurate, making it important to determine which transformation is most physically viable. Experiment determines
$T_c$ by identifying the temperature at which $\Delta \alpha$ vanishes \cite{Kai}. Subsequent work establishes that $\Delta\alpha=0$ only when the lattice parameters are
equal \cite{other4}, which is only consistent with the paraelectric phase having square P4/nmm symmetry [Fig.~\ref{fig:figure1}(b)] \cite{Mehboudi2016,other2}, and inconsistent
with hypotheses {\bf UOV 1} and {\bf UOV 2}.

\subsection{Understanding the effect of symmetry-induced constraints on energy barriers}\label{sec:iii.c}
Structural constraints induced by group symmetries make a transformation into the P4/nmm u.c.~preferable, and the next two paragraphs summarize the results along that direction that were presented in a companion article \cite{shiva}.

The glide operation in the Pnm2$_1$ group permits describing the u.c.~using two atoms only (say, those with coordinates $\mathbf{r}_3$ and $\mathbf{r}_2$), leading to atomic positions that are determined from eight variables (lattice parameters $a_1$ and $a_2$, and three atomic coordinates per atom). The lack of ferroelectric behavior along the $y-$direction implies that $\delta_y=0$ for both atoms $\mathbf{r}_2$ and $\mathbf{r}_3$, thus reducing the number of independent variables to six. Using the relative height of $\mathbf{r}_2$ and $\mathbf{r}_3$ instead of their individual heights, an explicit energy minimization in a space of five independent variational parameters leads to the u.c.~seen in Fig.~\ref{fig:figure1}(a) \cite{shiva}.

The square u.c.~with P4/nmm symmetry has two additional constraints: equal lattice parameters $a_1=a_2=a_s$ and $\delta_{x,s}=0$, so that the P4/nmm structure is reached by
an explicit energy optimization in a three-dimensional parameter space. The Pnmm structure has three added constraints: $a_1=a_{1,0}$, $a_2=a_{2,0}$, and $\delta_x=0$,
so that energy minimization in a space with only two independent variables is needed. In other words, the Pnmm structure is more constrained than P4/nmm, resulting in a
higher energy cost $J_r$ in realizing a Pnm2$_1 \to$ Pnmm  transformation when compared to the cost $J_s$ in the Pnm2$_1 \to$ P4/nmm transformation \cite{shiva}. Such
fundamentally different number of structural constraints should lead to larger critical temperatures in structures that retain the energetically disfavored rectangular u.c.~(so claims of ``numerical accuracy'' or ``limiting behaviors'' will have no bearing in accounting for dissimilar critical temperatures).

It is time to investigate the implications of these incommensurate theoretical descriptions on the structure, phonon spectra, and critical temperature(s) of these materials,
and to show that a topological Berezinskii-Kosterlitz-Thouless (BKT) transition fomented by rotational phonon modes corrects the physical picture of these transformations.

\subsection{Relevant soft optical vibrational modes for the two-dimensional structural transformation}\label{sec:iii.d}

Figure~\ref{fig:figure2} displays the vibrational spectra for a SnSe ML as obtained with the vdW-DF-cx functional. Its twelve eigenmodes are listed in ascending order at the
$\Gamma-$point in Fig.~\ref{fig:figure2}(a) (some modes swap order through the Brillouin zone and at the $M-$point).

The symmetries of phonon eigenmodes in Fig.~\ref{fig:figure2} are not affected by the choice of XC functional, and Tables \ref{ta:table1} and \ref{ta:table2} show the phonon
eigenfrequencies at the $\Gamma-$ and $M-$points, respectively, as obtained by the vdW-DF-cx and PBE XC functionals. The difference in listed frequencies according to XC
functional in these Tables for a given mode and $k-$point implies that interatomic forces depend on the choice of XC functional (because the electron density depends on the
XC approximation employed).

\begin{table}[tb]
\caption {Phonon eigenfrequencies $\nu$ (in THz) for a SnSe monolayer at the $\Gamma-$point for ground state (Pnm2$_1$), paraelectric square (P4/nmm), and paraelectric
rectangular (Pnmm) structures, according to the vdW-DF-cx and PBE XC functionals.}
\label{ta:table1}
\begin{center}
\begin{tabular}{|c||c|c|c||c|c|c|}
\hline
\hline
Phonon&vdW  &vdW &vdW     &PBE   &PBE   &PBE\\
mode  &Pnm2$_1$	&P4/nmm  &Pnmm   &Pnm2$_1$	&P4/nmm  &Pnmm\\
\hline
12	&5.75	&5.79	&5.90	&5.88	&5.90	&5.85\\
11	&4.71	&4.82	&4.91	&4.85	&4.92	&4.84\\
10	&3.63	&2.67	&2.51	&3.24	&2.65	&2.56\\
9	&3.18	&2.67	&2.48	&2.96	&2.65	&2.55\\
8	&2.82	&1.55	&1.70	&2.25	&1.80	&1.83\\
7	&2.07	&0.90	&0.93	&1.65	&1.06	&1.04\\
6	&1.72	&0.90	&0.04	&1.60	&1.06	&0.75\\
5	&1.44	&0.01	&$-$0.01&1.56	&0.00	&$-$0.01\\
4	&1.44	&0.01	&$-$0.04&1.43	&0.00	&$-$0.01\\
3	&0.00	&$-$0.01&$-$0.39&0.00	&0.00	&$-$0.02\\
2	&0.00&$-$1.31&$-$1.17&0.00	&$-$1.06&$-$0.99\\
1	&0.00&$-$1.31&$-$2.00&0.00	&$-$1.06&$-$1.51\\
\hline
\hline
\end{tabular}
\end{center}
\end{table}

\begin{table}[tb]
\caption {Phonon eigenfrequencies $\nu$ (in THz) for a SnSe monolayer at the $M-$point for ground state (Pnm2$_1$), paraelectric square (P4/nmm), and paraelectric rectangular
(Pnmm) structures, according to the vdW-DF-cx and PBE XC functionals .}
\label{ta:table2}
\begin{center}
\begin{tabular}{|c||c|c|c||c|c|c|}
\hline
\hline
Phonon&vdW  &vdW &vdW     &PBE   &PBE   &PBE\\
mode  &Pnm2$_1$	&P4/nmm  &Pnmm   &Pnm2$_1$	&P4/nmm  &Pnmm\\
\hline				
12	&5.02	&5.07	&5.17	&5.14	&5.17	&5.11\\
11	&5.02	&5.07	&5.16	&5.14	&5.17	&5.11\\
10	&3.41	&2.86	&2.75	&3.11	&2.80	&2.77\\
9	&3.41	&2.86	&2.75	&3.11	&2.80	&2.76\\
8	&2.36	&1.60	&1.65	&2.01	&1.63	&1.67\\
7	&2.36	&1.60	&1.62	&2.01	&1.63	&1.66\\
6	&2.12	&1.26	&1.23	&1.97	&1.32	&1.27\\
5	&2.12	&1.26	&1.21	&1.97	&1.32	&1.25\\
4	&1.50	&1.11	&0.67	&1.53	&1.25	&1.03\\
3	&1.50	&1.11	&0.60	&1.53	&1.25	&1.00\\
2	&1.10	&$-$0.55&$-$1.28&1.18	&0.28	&$-$0.75\\
1	&1.10	&$-$0.55&$-$1.32&1.18	&0.28	&$-$0.77\\
\hline
\hline
\end{tabular}
\end{center}
\end{table}

Returning to Fig.~\ref{fig:figure2}(a), one notes that modes 7 and 8 display quadratic dispersions at the $\Gamma-$point, and Fig.~\ref{fig:figure2}(b) shows modes 7 and 9
dispersing quadratically at the $M-$point as well (quadratic fittings and the associated parabolic coefficients are shown in red). According to Landau, {\em parabolic phonon modes can be rotational }\cite{Landau2}, and the consequences of this observation for the 2D transformations of these ferroelectrics will be made evident later on.

Figure \ref{fig:figure2}(c) displays the vibrational spectrum of the optimized P4/nmm structure. The phonon spectrum in Fig.~\ref{fig:figure2}(d) is for the optimized (though
energetically disfavored) Pnmm phase. The phonon dispersion of the ground state u.c.~seen in Fig.~\ref{fig:figure2}(a) appears in gray in these subfigures to assist in observing
the softening of multiple vibrational modes, especially at the $\Gamma-$ and $M-$points.

Phonon eigenvectors in Figs.~\ref{fig:figure2}(e) and ~\ref{fig:figure2}(f) reveal the relative reordering of relevant frequency eigenmodes upon structural transformations at the
$\Gamma-$ and $M-$ points, respectively. In the energetically favored P4/nmm structure, {\em two optical modes} 7 and 8 (vibrating along two orthogonal directions) soften the
most and are degenerate. This fact contradicts the single soft mode hypothesis {\bf UOV 3}.

Even in the Pnmm phase, these {\emph two} modes still soften the most, continuing to violate the {\emph single} soft mode assumption {\bf UOV 3}. The two degenerate soft optical modes at the $M-$point, which only vibrated unidirectionally at the $\Gamma-$point, now implicate vibrations along both $x-$ and $y-$directions, which is a further difficulty with {\bf UOV 3}. It is not granted that the model of Ref.~\cite{fei_prl_2016}, predicated on a single unidirectional optical mode at the $\Gamma-$point in an energetically disfavored u.c.~\cite{shiva} can confidently produce a quantitatively meaningful prediction of $T_c$.

\subsection{Evolution of order parameters and critical temperatures from finite-temperature calculations}\label{sec:iii.e}

We now turn to MD calculations to discover the thermal evolution of these MLs, and to make sense of their structural and vibrational properties discussed so far.

\subsubsection{Details of {\em ab initio} molecular dynamics calculations with the NPT and NVT ensembles}

{\em Ab initio} MD calculations with the NPT ensemble were carried out at a target pressure of 1 atm (or a minuscule 6$\times 10^{-5}$eV/\AA$^3$).
Atoms move at a finite $T$, creating non-zero forces and pushing containing walls along the way. The total force on atom $i$, $\mathbf{f}_{i,tot}$, is equal to the sum of
pairwise forces:
\begin{equation}
\mathbf{f}_{i,tot}=\sum_{\{n(i)\}}\mathbf{f}_{i,n(i)},
\end{equation}
where $\{n(i)\}$ is the set of neighbors to atom $i$ with a non-negligible pairwise force. These forces are used to make an initial guess for future atomic positions at each MD time step.

The stress tensor $\tensor{\sigma}$ is defined in terms of pairwise forces and distances among atoms subject to a pairwise force as follows:
\begin{equation}
\tensor{\sigma}=\frac{1}{V}\sum_{i=1}^N\sum_{n(i)}\mathbf{r}_{i,n(i)}\bigotimes \mathbf{f}_{i,n(i)},
\end{equation}
where $V$ is the supercell volume, $\mathbf{r}_{i,n(i)}$ is the relative position between the $i-$th atom and one of its neighbors $n(i)$, and $N$ is the number of atoms in
the supercell.

The crucial ingredient of a MD calculation within the NPT ensemble is that the containing walls are allowed to move to reduce built-up stress, such that the stress created by
finite-temperature atomic motion is no larger than the target pressure. In previous MD calculations performed on 8$\times$8 supercells, the accuracy of the critical temperature was assessed against multiple choices for barostats and thermostats and found no significant variations on its value against these empirical parameters \cite{other2}.

For direct comparison of our work with the UOV model, we carried out MD calculations within the NVT ensemble using {\em identical inputs} to the ones employed in MD runs
within the NPT ensemble.

One may be concerned that the MD results from the NPT and NVT ensembles should be identical at certain limiting situations. Indeed, there are two such limiting situations in
which results from these ensembles agree. They agree at $T=0$ K ({\em i.e.}, at a temperature irrelevant for the 2D structural transformation) as stress goes to zero in both
ensembles in the ground state.

\begin{figure}[tb]
\begin{center}
\includegraphics[width=0.48\textwidth]{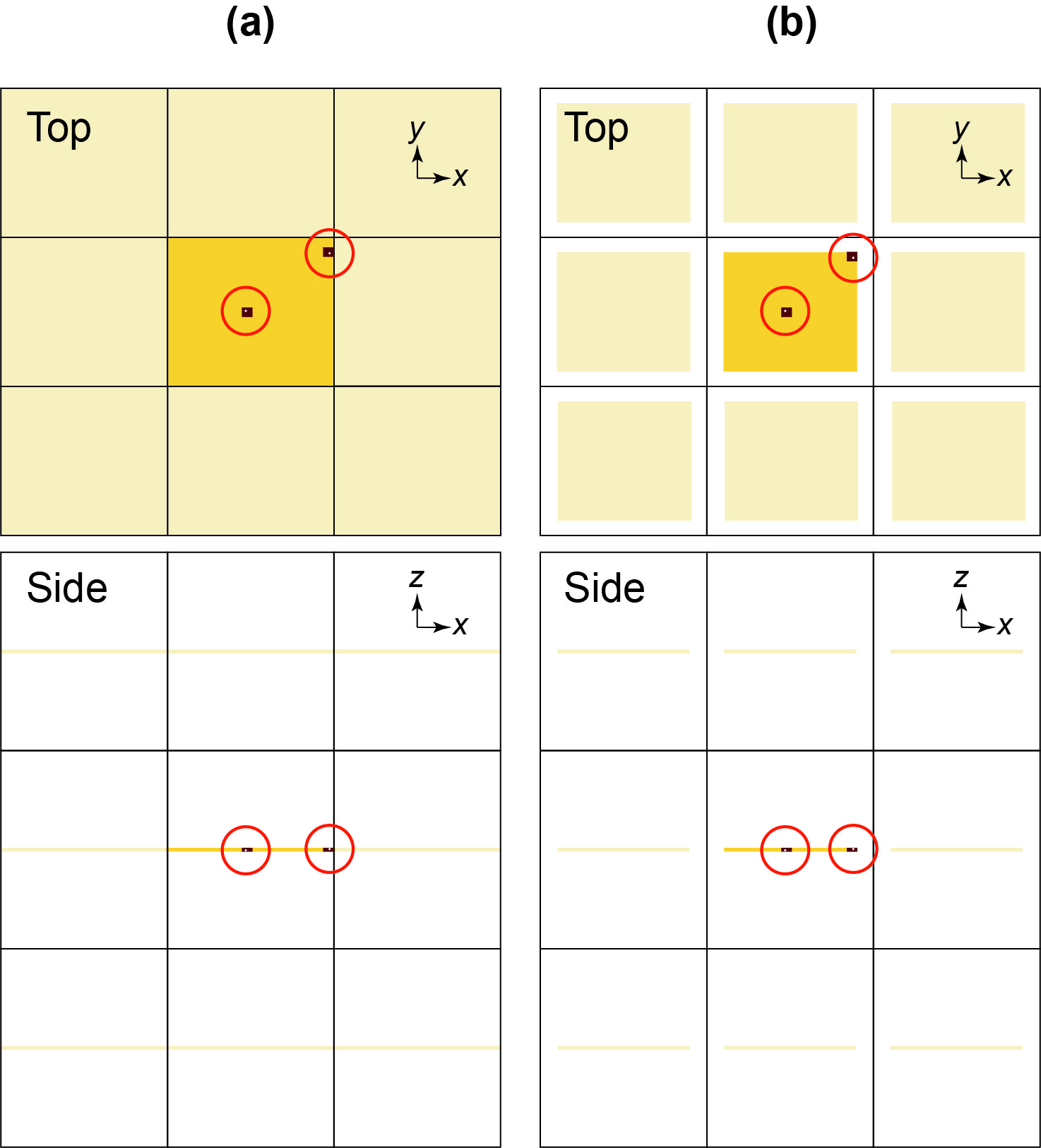}
\caption{(a) Top- and side-views of a 2D monolayer in a periodic cell calculation. The structure within the central supercell is shown in golden color, and atoms in
neighboring supercells are seen in a lighter color. The circles represent a sphere with a cutoff radius after which interatomic forces die off; there is sufficient vacuum to
disconnect periodic images along the out-of-plane ($z$) direction. (b) Top- and side-views of a 2D monolayer with vacuum around the edges of the finite-size
flake.}\label{fig:figure3}
\end{center}
\end{figure}

The second limiting situation occurs as follows: As indicated before, at finite $T$ and when atoms are near the containing walls such as in periodic calculations, atomic
motion exerts pressure onto such containing walls [shown by solid black lines in Fig.~\ref{fig:figure3}(a)] and pushes them away from their zero-temperature magnitudes until a
target pressure is reached in the NPT ensemble. These walls remain fixed to their zero-temperature magnitudes when employing the NVT ensemble, nevertheless, building a
temperature-dependent stress $\tensor{\sigma}$.

\begin{figure*}[tb]
\begin{center}
\includegraphics[width=0.96\textwidth]{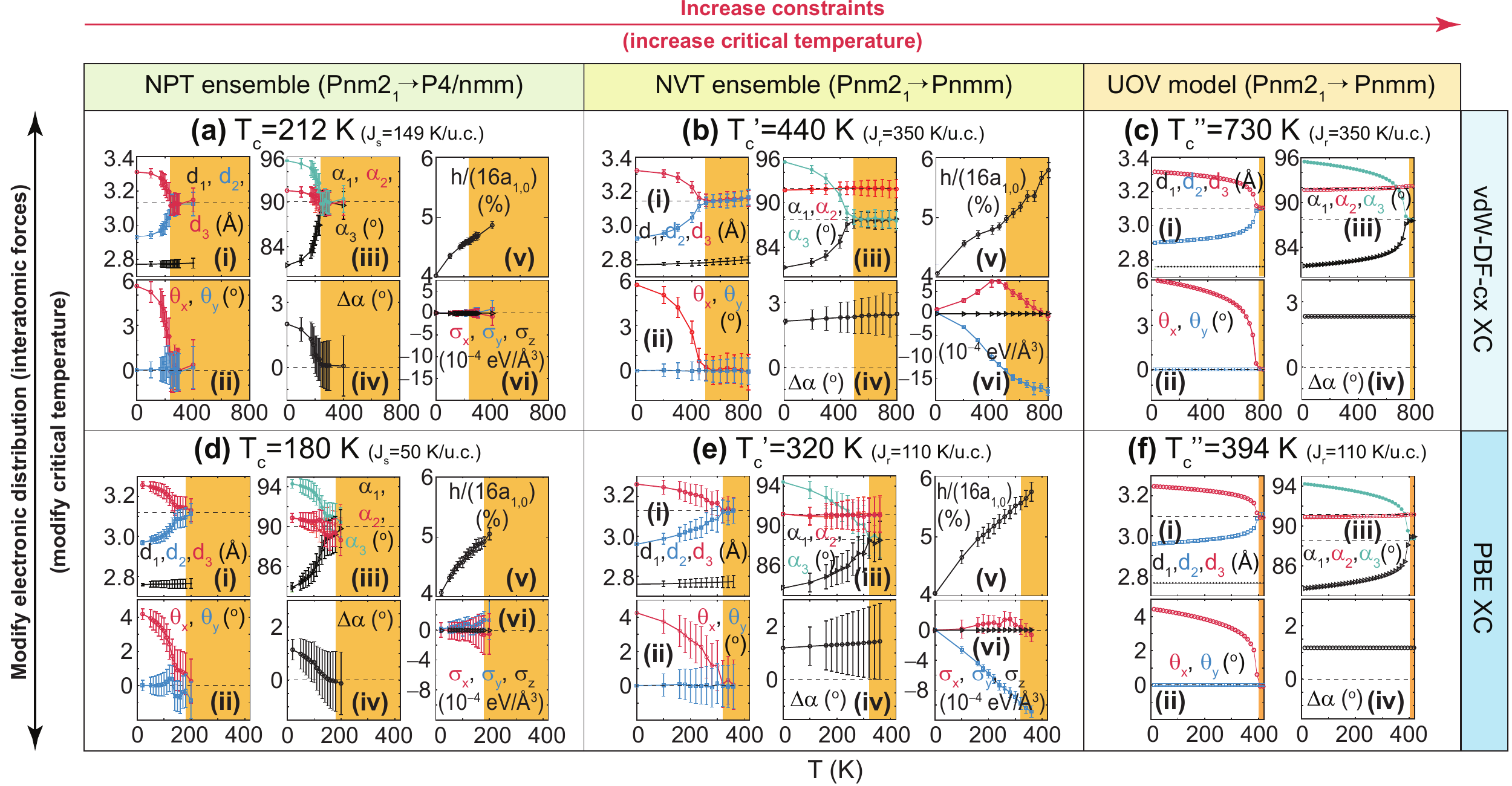}
\caption{Critical temperatures for a SnSe ML according to different ensembles/models (columns) and two XC functionals (rows). Once an ensemble/model and XC functional are
chosen (separate panels), the thermal evolution of (i) $d_1$, $d_2$ and $d_3$, (ii) $\theta_x$ and $\theta_y$, (iii) $\alpha_1$, $\alpha_2$ and $\alpha_3$, (iv) $\Delta
\alpha$, (v) $h$, and (vi) $\sigma$ are listed. Orange boxes indicate $T$ at which the ferroelectric to paraelectric 2D transformation has been reached. $T_c$, $T_c'$ and
$T_c''$ as well as $J_s$ and $J_r$ are strongly dependent on XC functionals, {\em and increase as structural constraints are added}.}
\label{fig:figure4}
\end{center}
\end{figure*}

But an alternative situation is to include a vacuum between the flake and the containing walls, as depicted in Fig.~\ref{fig:figure3}(b). The periodic images will not contribute
to the stress $\tensor{\sigma}$ when the vacuum is thicker than the effective neighbor cutoff radius, schematically shown by red circumferences in the Figure. (In our
calculations, this is already the case along the $z-$direction since these are 2D materials, and indeed both ensembles will agree that there is zero stress along that
direction on Fig.~\ref{fig:figure4} later on.) Even in the NVT ensemble then, though the volume of the containing walls does not change, the lattice parameters of the flake can
freely evolve, making $\Delta\alpha=0$ at $T_c$. {\em Only in that context do the NPT and NVT ensembles at finite $T$ agree in the thermodynamic limit}. In any event, it is
inappropriate to fix the volume from the outset when one expects significant changes to lattice vectors.

In the context of this study, we employed the NVT ensemble with boundary conditions as shown in Fig.~\ref{fig:figure4}(a) with the explicit goal of showing what happens when one retains fixed lattice parameters and to demonstrate its
fundamental differences with respect to the UOV model.

\subsubsection{Comparative thermal evolution of order parameters and stress for the NPT and NVT ensembles and the UOV model}

Full details of the parametrization of the UOV model can be found in Appendix \ref{sec:appendixA}, and Fig.~\ref{fig:figure4} displays the thermal evolution of the structural order parameters introduced in Fig.~\ref{fig:figure1}. In each panel, the first, second, and third nearest neighbor distances ($d_1$, $d_2$ and $d_3$) are displayed in subplot (i). Subplot (ii) contains the angle $\theta_x$ introduced in Ref.~\cite{fei_prl_2016}. Subplot (iii) displays the angles $\alpha_1$, $\alpha_2$ and $\alpha_3$, and subplot (iv) shows the rhombic distortion angle $\Delta\alpha$ \cite{Kai,other4}. In Figs.~\ref{fig:figure4}(a,b,d,e), subplot (v) displays the calculation cell height of these materials, and subplot (vi) shows the diagonal components of the stress tensor. These last two mentioned panels are not shown for the UOV model \cite{fei_prl_2016} because it enforces a constant $h$ and has no prescription for evaluating $\tensor{\sigma}$. The orange rectangles in Fig.~\ref{fig:figure4} indicate the critical temperatures at which the neighbor distances $d_2$ and $d_3$ coalesce [subplot (i)] and the angle $\theta_x$ is extinguished [subplot (ii)]. These two features are present in each of the finite-$T$ calculations considered.
However, the angles $\alpha_1$, $\alpha_2$ and $\alpha_3$ [subplot (iii)] reveal the different symmetry of the resultant paraelectric phases. For the NPT ensemble these three
angles coalesce to $90^{\circ}$ and the structure has P4/nmm symmetry. For the NVT ensemble and UOV model in which lattice parameters are not allowed to evolve with $T$,
their values converge to two $\alpha_1<90^{\circ}<\alpha_2$, and the structure has twofold Pnmm symmetry. Both the NVT ensemble and the UOV model \cite{fei_prl_2016} display
$\Delta\alpha>0$ for all $T$, meaning that these two models do not reproduce the collapse of $\Delta\alpha$ observed in experiment \cite{Kai}.

These ferroelectric materials swell in going from the layered bulk to a ML \cite{shiva}, implying that the weak bonds keeping layers apart are responsible for the structure
within isolated MLs, too. These weak bonds owing to lone pair electrons are better treated with dispersive corrections, and so we performed calculations with the vdW-DF-cx XC
functional. We also used the PBE functional [(choice {\bf UOV 6}) Figs.~\ref{fig:figure4}(d-f)] to be able to compare directly with the assumptions and numerical results from
the UOV model \cite{fei_prl_2016}. At the moment, it is unclear which XC functional better describes the electronic density of these peculiar 2D ferroelectrics.

At finite $T$, the NPT ensemble equilibrates the pressure at every step, eliminating the in-plane stresses [Fig.~\ref{fig:figure4}(a,d)]. But for the NVT ensemble,
the in-plane stresses ($\sigma_x$ and $\sigma_y$) are not allowed to relax; so $\sigma_x$ and $\sigma_y$  take on large magnitudes while the component of stress along the
vertical direction ($\sigma_z$) remains zero [Fig.~\ref{fig:figure4}(b,e)]. The negative magnitude of $\sigma_y$ implies that the ML presses against its containing walls along
the $y-$direction, or that the lattice parameter along this direction would increase its magnitude if it were not constrained by the containing wall. The positive magnitude
of $\sigma_x$ increases for $0<T<T_c'$ and implies that the ML pulls on its containing walls along the $x-$direction, and such pull is alleviated to some extent by the
transformation to the Pnmm phase. At temperatures above $T_c'$, $\sigma_x$ decreases monotonically in the temperature range examined here.

The UOV model, like the NVT ensemble in which the atoms exert non-negligible stress on periodic images, is based on lattice parameters pinned to their zero temperature values
and builds stress in a similar manner (though it has no formal manner to quantify such stress), however, it cannot be reconciled with the experimentally-relevant NPT result
(which recovers the fourfold symmetry). The important point to recognize is that, rather than the $\Delta\alpha>0$ predicted by the UOV model in Ref.~\cite{fei_prl_2016} [subplots (iv) in Figs.~\ref{fig:figure6}(c) and \ref{fig:figure6}(f)],
both MD approaches can be made consistent with $\Delta\alpha=0$ at $T_c$ as in experiment and in the simple Potts model \cite{Mehboudi2016}.

A straightforward way to make sense of the increased critical temperatures in going from the far left to the far right in Fig.~\ref{fig:figure4} is by understanding that the
gradual structural constraints being imposed (constant area, reduced degenerate ground states, suppressed active soft modes, constrained vibrational direction, covariant angle approximation, and so on) are generalizations of the strain constraint introduced in Refs.~\cite{fei_prl_2016} and \cite{other4}.

One should also rely on classical work on phase transformations in 2D to gain additional insight. The number of degenerate ground states in Potts model is labeled $r$. Using all four degenerate ground states \cite{Mehboudi2016,otherarticle,other3} ($r=4$), he writes $2J/k_BT_c=1.76$. He also says that $2J/k_BT_c=0.88$ for $r=2$ ({\bf OUV 1}) \cite{potts}. So even if the energy barriers $J$ were not to change in the two calculations, Potts estimates a twice as large critical temperature if two degenerate ground states are used \cite{fei_prl_2016} instead of four. This analytical result emphasizes the fact that the critical temperature is expected to increase as soon the first {\bf OUV 1} hypothesis is enforced. No limiting behavior connects a model based on two degenerate ground states and another based on four.

\subsection{Failure of the angle-covariant approximation}\label{sec:iii.f}

\begin{figure}[tb]
\begin{center}
\includegraphics[width=0.48\textwidth]{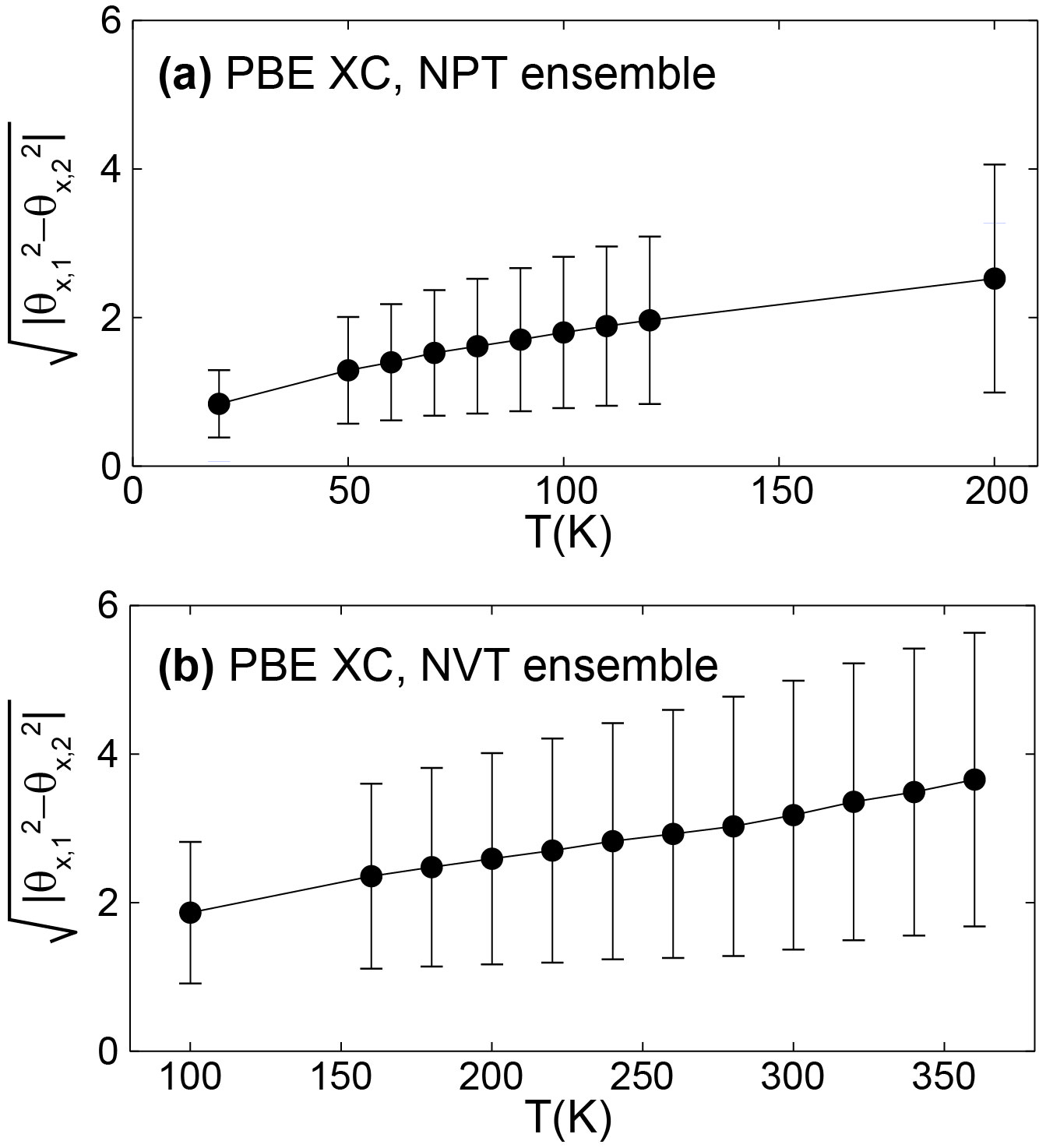}
\caption{Violation of the angle-covariant approximation in {\em ab initio} molecular dynamics calculations, irrespective of ensemble. The angle-covariant approximation holds
if $\sqrt{|\theta_{x,1}^2-\theta_{x,2}^2|}=0$.}\label{fig:figure5}
\end{center}
\end{figure}

The angle-covariant approximation (assumption {\bf UOV 4}) can be tested by recording the difference between the absolute values of $\theta_{x,1}$ and $\theta_{x,2}$ within a
given unit cell, and Fig.~\ref{fig:figure5} displays the time and unit cell average of $\sqrt{|\theta_{x,1}^2-\theta_{x,2}^2|}$, where the values of  $\theta_{x,1}$ and
$\theta_{x,2}$ are consistently extracted from the same unit cell. The angle-covariant approximation ($\theta_{x,1}=\theta_{x,2}$ at each unit cell) holds when
$\sqrt{|\theta_{x,1}^2-\theta_{x,2}^2|}=0$. Clearly, these 2D materials fail to obey such assumption, too.

\subsection{Breakdown of the unidirectional optical vibration approximation}\label{sec:iii.g}

\begin{figure}[tb]
\begin{center}
\includegraphics[width=0.48\textwidth]{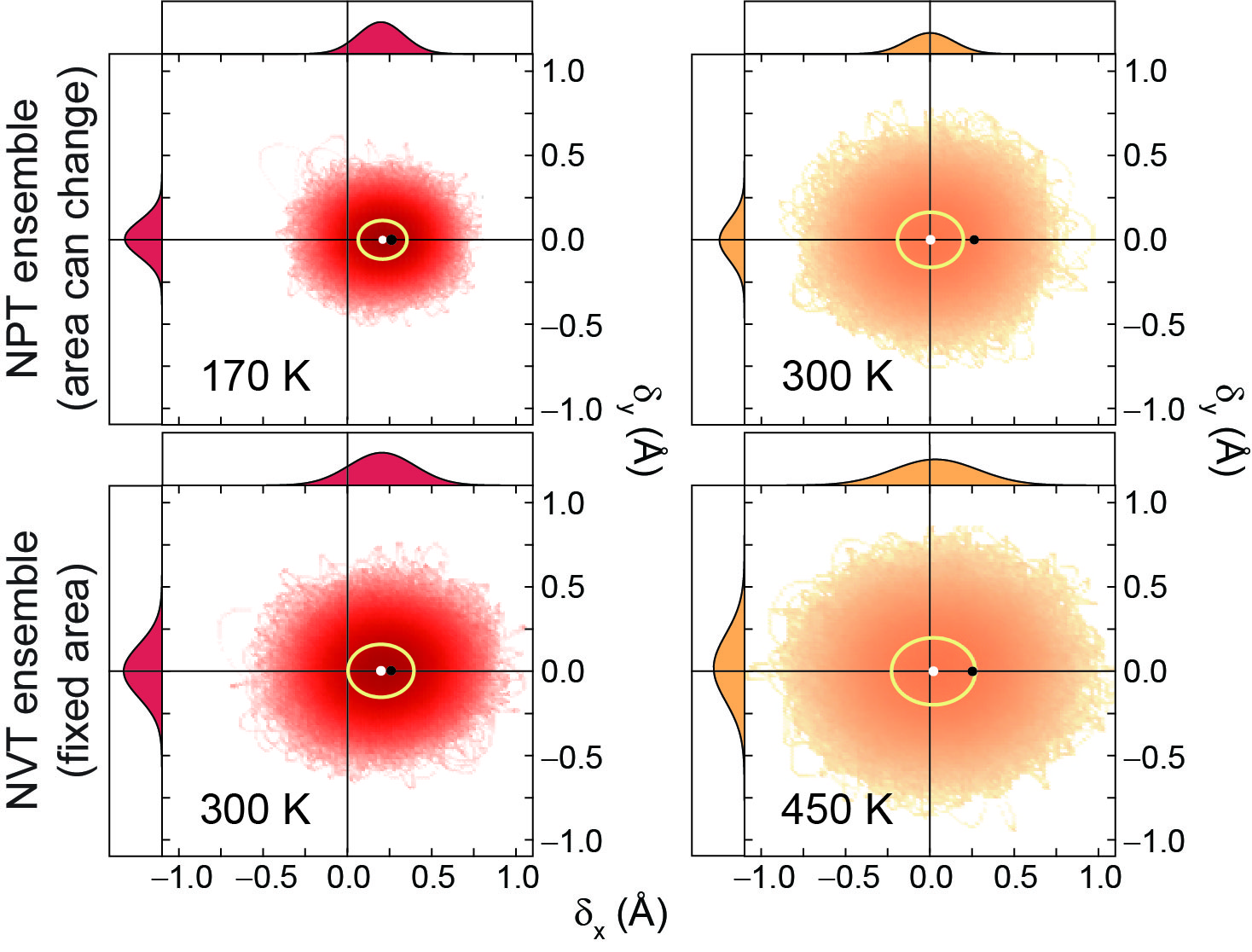}
\caption{Distribution of $\delta_x$ and $\delta_y$ at finite $T$ within the NPT and NVT ensembles. The black dot indicates $\delta_{x,0}$, and the white dot (yellow curve)
the average (standard deviation) of ($\delta_x,\delta_y$) for $T$ as indicated; $\delta_x$ coalesces to zero at a smaller $T$ on the NPT ensemble. The near radial symmetry of
$\delta_x$ and $\delta_y$ at finite $T$ evidences rotational motion.}\label{fig:figure6}
\end{center}
\end{figure}

The rotons predicted in Fig.~\ref{fig:figure2}(b) are in-plane rotations of the nearest-neighbor bonds depicted in orange in Fig.~\ref{fig:figure1}(a). The distributions of
instantaneous projections of these bonds along the $x-$ and $y-$directions at finite $T$ ($\delta_x$ and $\delta_y$) are seen in Fig.~\ref{fig:figure6} above and below $T_c$
($T_c'$). The near symmetry of the distributions of $\delta_x$ and $\delta_y$ implies rotations underpinned by the soft phonon modes emphasized with yellow and orange boxes
in Fig.~\ref{fig:figure2}(e,f). The fixed volume constraint in the NVT ensemble does not preclude rotational motion, and this differentiates it from the UOV model in which
(i) dynamical oscillations of modes eight and nine were discarded, and (ii) the evolution of mode seven on the full Brillouin zone was not accounted for. In a modification to
reproduce our results, the UOV model should also include a prescription to release stress.

Based on the discussion and results from Sec.~\ref{sec:iii.a} to ~\ref{sec:iii.f}, there can be no expectation that the critical temperature as obtained from {\em ab initio}
MD agrees with the estimate provided by the UOV model when there are so many fundamental assumptions that go unpreserved on the latter.

\begin{figure*}[htb!]
\begin{center}
\includegraphics[width=0.96\textwidth]{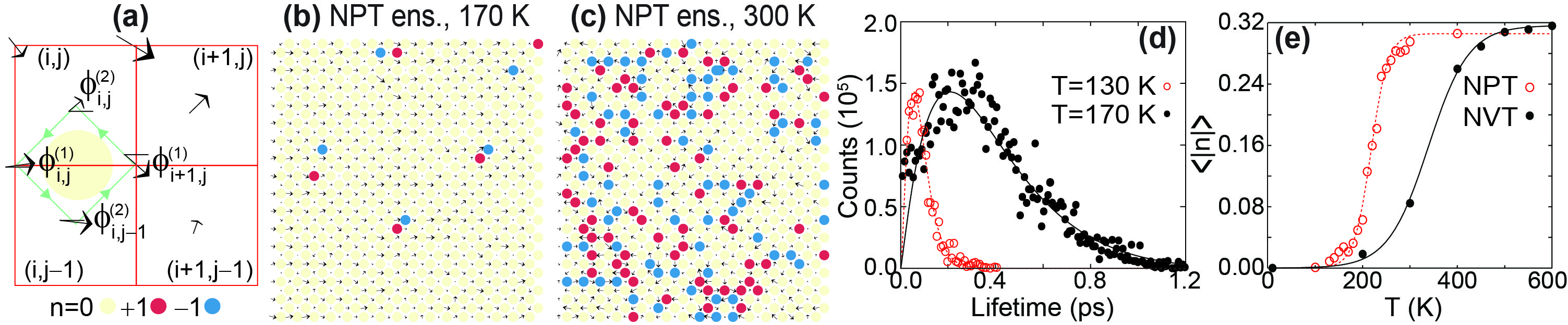}
\caption{(a) Vorticities are computed within four-point closed loops. (b) Snapshot of the projections $(\delta_x,\delta_y)$ at a given time $t$ for a calculation within the
NPT ensemble at $T<T_c$. (c) Similar plot for $T>T_c$. (d) Distribution of vortex lifetimes for the NPT ensemble. (e) The number of vortices increases at $T_c$ ($T_c'$) and
eventually saturates in the NPT (NVT) ensemble.}\label{fig:figure7}
\end{center}
\end{figure*}

\subsection{The non-trivial topology of the two-dimensional structural transformation}\label{sec:iii.h}

Rotations owing to the softened phonon modes lead to intriguing vortex physics. In the BKT transition, the Mermin-Wagner-Hohenberg prohibition on long-range order in a 2D
system of planar rotators is avoided by the existence of topological vortices. This was impressively successful at describing the superfluid transition physics of 2D $^4$He
films \cite{BR1}. At and above the transition temperature the existence of vortices can lower the
free energy via the dominating entropy term. Vortex and anti-vortex pairs are tightly bound to each other and do not disrupt the long-range orientation below $T_c$. Above
$T_c$, the pairs unbind and vortices/antivortices are free to exist throughout the system. The rotations hinted at in Fig.~\ref{fig:figure2}(b) and Fig.~\ref{fig:figure6} create
non-zero winding numbers $n$ in the BKT circulation integral \cite{KostThou1973,KostRev2016}.

Vortices/antivortices are identified by their winding numbers, computed by a four-point sum \cite{Nahas2017} along the counterclockwise path shown in green in
Fig.~\ref{fig:figure7}(a). For every time step, we track the azimuthal angle $\phi(i,j)$ of the vector $(\delta_x,\delta_y)$ in each unit cell $(i,j)$, which takes values in
$(-\pi,\pi$]. Then the winding number is given by
\begin{eqnarray}\label{eq:e2}
&2\pi n = [\phi_{2}(i,j)-\phi_{1}(i+1,j)]+[\phi_{1}(i,j)-\phi_{2}(i,j)]+ \nonumber\\
&[\phi_{2}(i,j-1)-\phi_{1}(i,j)]+[\phi_{1}(i+1,j)-\phi_{2}(i,j-1)].
\end{eqnarray}
Here the time dependence is obviated, and the $[\phi-\phi']$ notation indicates that the difference of the angles is taken as the principal value in the range $(-\pi,\pi$].
Antivortices ($n=-1$), vortices ($n=+1$), and other ($n=0$) dual lattice sites are indicated as blue, red, and yellow circles, respectively, in Fig.~\ref{fig:figure7}(b,c).

To calculate the vortex lifetime shown in Fig.~\ref{fig:figure7}(d), we probed whether a given lattice site or its nearest neighbors are occupied by a vortex of one charge at every timestep. The time counter resets whenever this condition is not met. Though we have a time resolution of 1.5 fs, we present the distribution of lifetimes using a bin size of 9 fs for clarity. The solid lines are fits to Gamma distributions. At T=130 K, the average lifetime is 85 fs, and at T=170 K the average lifetime is 390 fs. Vortices persist for many tens of timesteps (or hundreds for higher temperature) near their starting position and can be called robust in that sense. Fig.~\ref{fig:figure7}(e) indicates that the number of vortex-antivortex pairs increases and saturates with $T$ and that their mean number at a given $T$ depends on the ensemble being employed.


\section{Conclusions}\label{sec:conclusions}

In summary, this work differentiates between two possible structural transformations in group-IV monochalcogenide MLs at finite temperature by a direct comparison of their assumptions. Phonon modes indicate that a hypothesis of unidirectional vibrations may not be justified; the importance of rotations is hinted at by the existence of quadratically dispersing roton modes in the phonon dispersion. The insight gained in these unit cell calculations is put to the test in MD calculations, and the P4/nmm u.c.~was proven to have a lower critical temperature than the Pnmm structure under otherwise identical conditions. Even the Pnmm structure is reached at a lower critical temperature when employing the NVT ensemble than within the UOV model, because none of the rotations are artificially discarded in the former. This way, differences on critical temperatures are fundamentally due to an increase of structural and/or dynamical constraints; the lowest critical temperature will be the most viable from a physical standpoint. The choice of XC functional clearly has a significant effect on estimating the critical temperature, and determination of the optimal functional will have to be informed by further experiment. Finally, the rotational modes were shown to give rise to a topological structural transformation.

We remain unaware of any report analyzing such intriguing structural transformation with the depth of focus hereby provided, and believe that these results can help galvanize
the field of 2D ferroelectrics and make it move forward with renewed confidence. Effects of domain walls on $T_c$ is an interesting topic that should ought to consider
domain-size effects and therefore lies beyond this work's scope.

\begin{acknowledgments}
We acknowledge conversations with K.~Chang and L.~Bellaiche and thank A.~Pandit for assistance. J.W.V.~and S.B.L.~were funded by an Early Career Grant from the U.S.~DOE
(DE-SC0016139). Calculations were performed on Cori at NERSC, a U.S.~DOE Office of Science User Facility (DE-AC02-05CH11231).
Pinnacle supercomputers, funded by the NSF, the Arkansas Economic Development Commission, and the Office of the Vice Provost for Research and Innovation.
\end{acknowledgments}

\appendix
\section{Details of UOV calculations}\label{sec:appendixA}

\begin{figure}[tb]
\begin{center}
\includegraphics[width=0.48\textwidth]{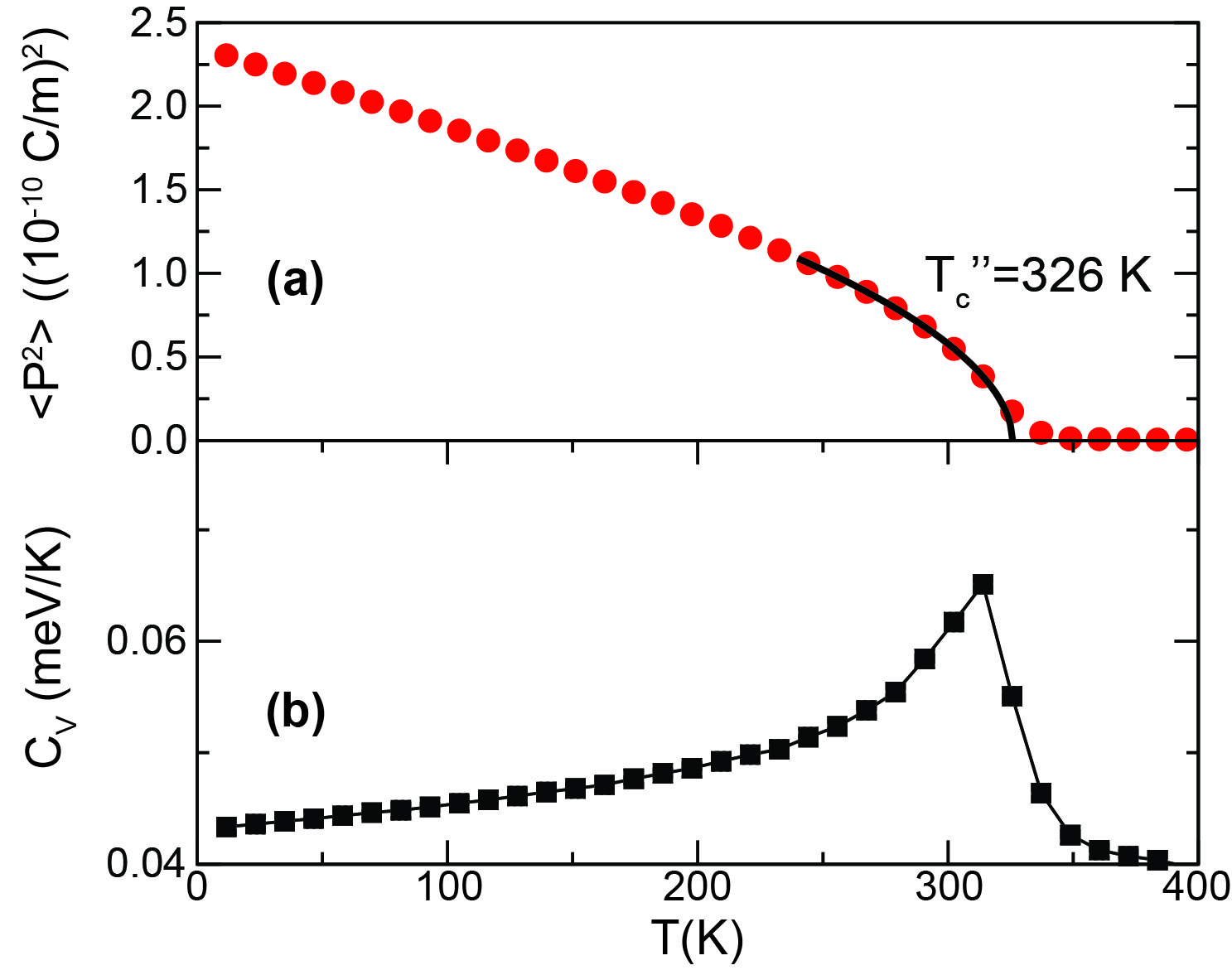}
\caption{(a) Thermal evolution of the intrinsic dipole moment $P$, using the parameters provided in Ref.~\cite{fei_prl_2016} based on the PBE XC functional (choice {\bf UOV
5}). (b) Specific heat as obtained with that model. $T_c$ agrees with the value reported in that work when using their fitting parameters, thus validating Monte Carlo results
to be presented later on.}\label{fig:figure8}
\end{center}
\end{figure}

Figure \ref{fig:figure8} verifies the accuracy of the Monte Carlo solver, as it reproduces the critical temperature $T_c''$ predicted in the UOV model when using their fitting
parameters \cite{fei_prl_2016}.

Unlike the energy landscape created as a function of lattice parameters $a_1$ and $a_2$ in Refs.~\cite{Mehboudi2016,other3,other4}, the UOV model \cite{fei_prl_2016} fixes
$a_1$ and $a_2$ to their zero-temperature values, and proposes a landscape as a function of the two unit cell tilt angles $\theta_{1,x}$ and $\theta_{2,x}$ in
Fig.~\ref{fig:figure1}(a) (hypotheses {\bf UOV 1} and {\bf UOV 2}).

\begin{table}[tb]
\caption {Fitting parameters for the UOV model in Eqs.~(\ref{eq:landscape1}), (\ref{eq:landscape2}) and (\ref{eq:interaction}).}
\label{ta:table4}
\begin{center}
\begin{tabular}{|c||c|c|}
\hline
\hline
Parameter & vdW-DF-cx XC & PBE XC\\
\hline
$A_1$ ($\times 10^4$ K/(rad$^2$u.c.))&  $-$3.453   &    $-$2.032 \\
$A_2$ ($\times 10^4$ K/(rad$^2$u.c.))&  $-$0.255   &    1.277    \\
$B_1$ ($\times 10^4$ K/(rad$^4$u.c.))&  84.426     &    92.502   \\
$B_2$ ($\times 10^4$ K/(rad$^4$u.c.))&  53.767     &    55.360   \\
$B_3$ ($\times 10^4$ K/(rad$^4$u.c.))&  607.254    &   597.710   \\
$C_1$ ($\times 10^4$ K/(rad$^6$u.c.))& $-$5035.864 & $-$5510.669 \\
$C_2$ ($\times 10^4$ K/(rad$^6$u.c.))& $-$4655.471 & $-$5944.196 \\
$C_3$ ($\times 10^4$ K/(rad$^6$u.c.))& $-$203.185  &  $-$460.243 \\
$F_1$ ($\times 10^4$ K/(rad$^2$u.c.))& 3.544       &  4.152      \\
\hline
\hline
\end{tabular}
\end{center}
\end{table}

\begin{figure}[tb]
\begin{center}
\includegraphics[width=0.48\textwidth]{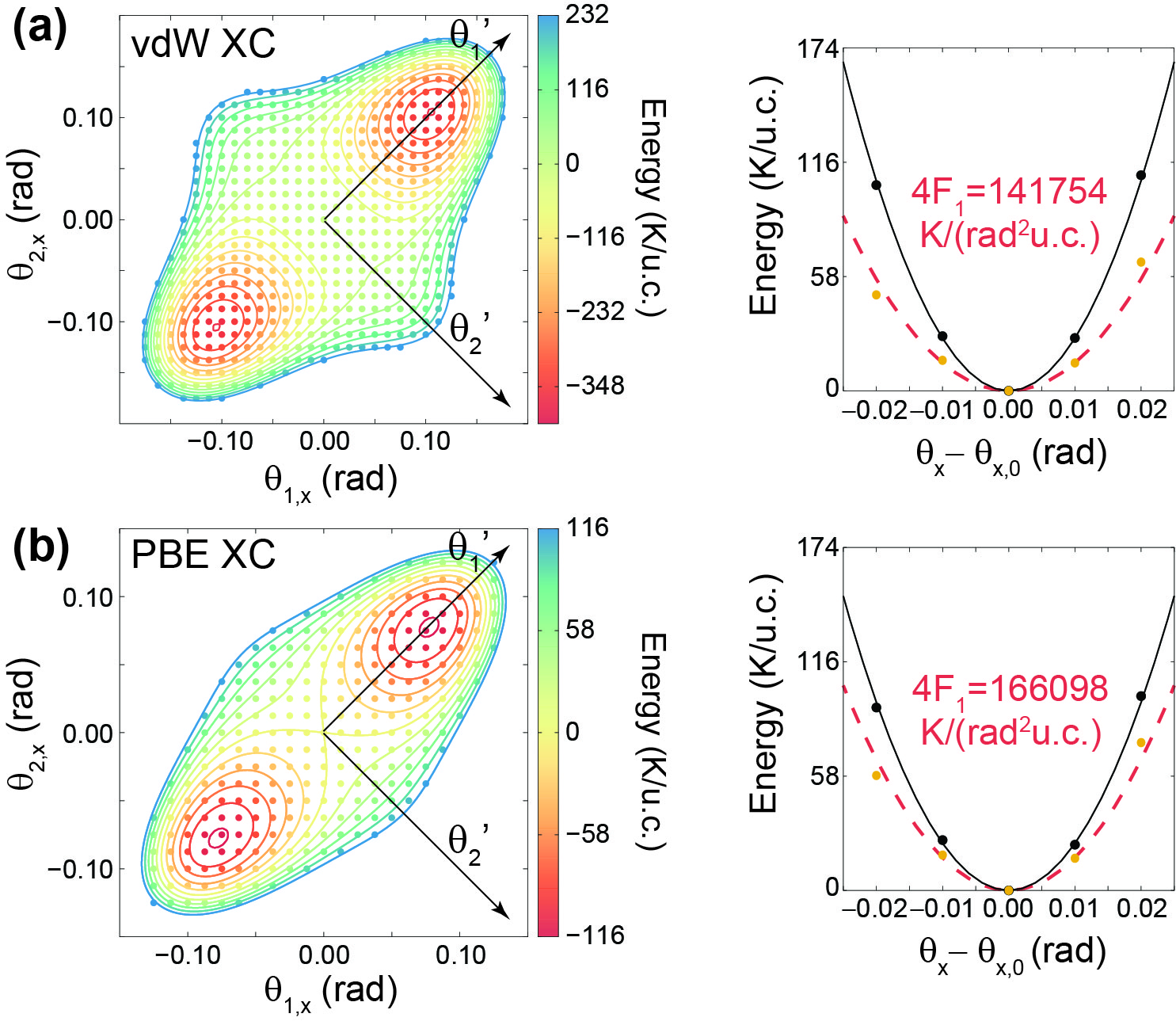}
\caption{Energy landscape as a function of $\theta_{1,x}$ and $\theta_{2,x}$ in the rectangular structure with zero-temperature lattice parameters $a_{1,x}$ and $a_{2,x}$,
using the (a) vdW-DF-cx and (b) PBE XC functionals. Plots to the right indicate the mean field coupling of neighboring dipoles without subtracting a strain contribution
(solid black) or when subtracting it (dashed red curve). Use of the solid black curve would increase $T_c''$ even further, and the red curve was employed in computing
$T_c''$.}\label{fig:figure9}
\end{center}
\end{figure}

In principle, there are two values for the tilting angle, $\theta_{1,x}$ and $\theta_{2,x}$. The authors of the UOV model set $\theta_{1,y}$ and $\theta_{2,y}$ equal to zero,
to enforce unidirectional vibrations, effectively freezing phonons oscillating along the $y-$direction (hypothesis {\bf UOV 2}). Energy landscapes are shown in
Fig.~\ref{fig:figure9} as a function of $\theta_{1,x}$ and $\theta_{2,x}$ for our two choices of XC functionals.

\begin{figure}[tb]
\begin{center}
\includegraphics[width=0.48\textwidth]{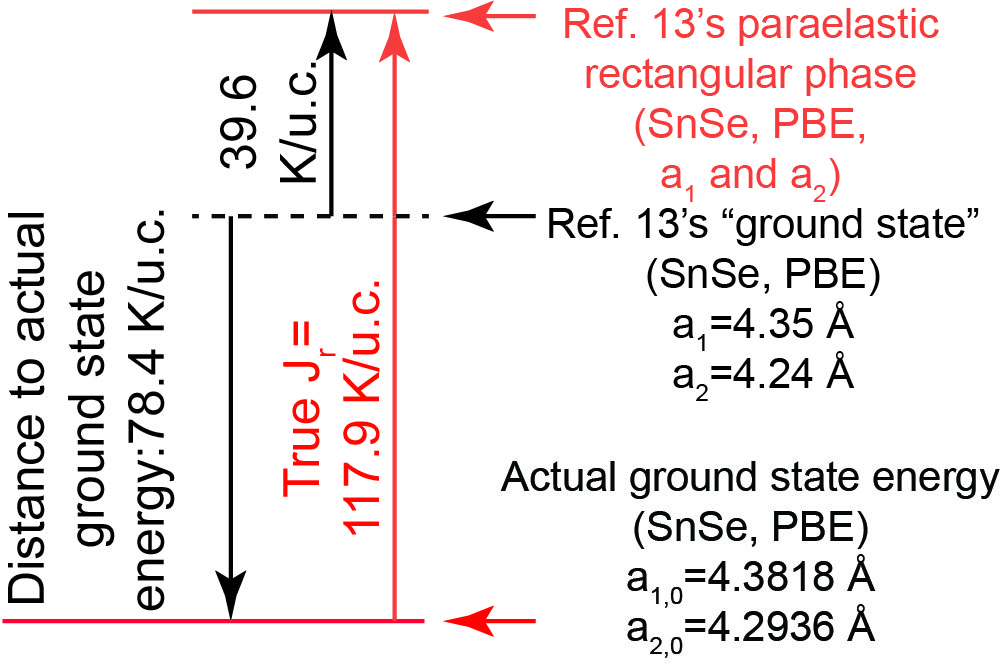}
\caption{$J_r$ is underestimated in Ref.~\cite{fei_prl_2016} (they report 39.6 K/u.c. while its actual value is 117.9 K/u.c.), as determined using exactly the same numerical
tool but well-converged inputs. See Ref.~\cite{shiva} for more details, and Ref.~\cite{JAP} for another work with lattice parameters in exact agreement with ours. This numerical inaccuracy impacts the magnitude of $T_c''$ estimated in Ref.~\cite{fei_prl_2016}.}\label{fig:figure10}
\end{center}
\end{figure}

The transformation $\theta_1'=\frac{\theta_{1,x} + \theta_{2,x}}{\sqrt{2}}$ and $\theta_2'=\frac{\theta_{1,x} - \theta_{2,x}}{\sqrt{2}}$ indicated in Ref.~\cite{fei_prl_2016}
permits writing an expression for the energy landscape as:
\begin{eqnarray}\label{eq:landscape1}
&E(a_{1,0},a_{2,0},\theta_1',\theta_2') = A_1\theta_1'^2 + A_2 \theta_2'^2\\
&+ B_1 \theta_1'^4 + B_2 \theta_2'^4 + B_3 (\theta_1'\theta_2')^2\nonumber\\
&+ C_1 (\theta_1'\theta_2'^2)^2 + C_2 (\theta_1'^2\theta_2')^2 + C_3 \theta_1'^6,\nonumber
\end{eqnarray}
with the use of zero-temperature lattice parameters made explicit.
The continuous, colored equal energy contours in Fig.~\ref{fig:figure9}(a,b) demonstrate how closely the fit hews to the landscape which was calculated on a regular mesh of
points. The fitting coefficients are in Table \ref{ta:table4}. (A hypothetical $\theta_2'^6$-term, admissible by symmetry, had a nonsignificant fitting coefficient.) The
lowest energy values seen on the landscapes are $J_r=350$ K/u.c. when using the vdW XC functional, and 110 K/u.c. when the PBE XC functional is employed. Slightly different
numerical values when employing the PBE ensemble in between this work and Ref.~\cite{fei_prl_2016} are due to the use of different computational codes, but the main points of
this work remain unchanged by these numerical considerations. In addition, and as shown in Fig.~\ref{fig:figure10} and Ref.~\cite{shiva}, there are reasons to believe the numerical fitting parameters in Ref.~\cite{fei_prl_2016} could be improved. Being more explicit, the barrier they report turns out to be about 40 K/u.c. We reproduce their prediction when using their lattice parameters. Nevertheless, multiple works now show their inaccuracy \cite{shiva,JAP}. As seen in Fig.~\ref{fig:figure10}, the accurate ground state lattice parameters yield a barrier that is three times larger than the one estimated in Ref.~\cite{fei_prl_2016}: since all their parameters were created using incorrect lattice vectors, their $T_c''$ (326 K) gets to be lower than the correct one within their own model ($\sim$ 400 K), as reported in Fig.~\ref{fig:figure4}(f).

The further constraint on that model is setting $\theta_2'=0$ to reduce the 2D landscape onto its diagonal. It implies $\theta_{1,x}=\theta_{2,x}\equiv \theta_x$. This {\em
angle-covariant approximation} (assumption {\bf UOV 4}) leads to a one-dimensional double-well landscape:
\begin{eqnarray}\label{eq:landscape2}
&E(\theta_x) = \sum_{i,j} (2 A_1 \theta_{x}(i,j)^2 + 4 B_1 \theta_{x}(i,j)^4\\
& + 8 C_3 \theta_{x}(i,j)^6 + {{\cal H}_{<i,j;i',j'>}}),\nonumber
\end{eqnarray}
with $i$ and $j$ representing u.c. along the $x-$ and $y-$directions, respectively. As seen in Fig.~\ref{fig:figure5}, assumption {\bf UOV 4} does not hold in realistic
situations.

The first three terms to the right represent an on-site energy, while the fourth one is an interaction term arising from different angle-covariant orientations in the four
nearest neighbor u.c.:
\begin{eqnarray}\label{eq:interaction}
&{{\cal H}_{<i,j;i',j'>}}=\\
&F_1\{(\theta_{x}(i,j) - \theta_{x}(i+1,j))^2+(\theta_{x}(i,j)-\theta_{x}(i-1,j))^2 \nonumber\\
&+(\theta_{x}(i,j) - \theta_{x}(i,j+1))^2+(\theta_{x}(i,j)-\theta_{x}(i,j-1))^2 \}\nonumber,
\end{eqnarray}
with terms of the form $(\theta_{y}(i,j) - \theta_{y}(i',j'))^2$ and $(\theta_{x}(i,j) - \theta_{y}(i',j'))^2$ implicitly excluded, amounting to the freezing of optical
phonon modes along the $y-$direction and the avoidance of possible correlated motion along the $x-$ and $y-$directions.

$F_1$ was obtained the following way: the value of $\theta_{x,0}=\text{arcsin}(\delta_{x,0}/d_{1,0})$ corresponding to the ground state structure was set onto an $8\times 8$
supercell. The magnitude of $\theta_x$ in a single, ``central'' unit cell was left to vary afterwards leading to the change in energy displayed by black circles (i.e., strict
angle-covariance is enforced), and the energy exhibits the parabolic dispersion on the solid black line seen on the plots to the right in Fig.~\ref{fig:figure9}. This is the mean-field approximation (assumption {\bf UOV 5}), that is absent in MD calculations. Considering that
the angle-covariant tilt away from $\theta_{x,0}$ in the ``central'' unit cell produces on-site stress as well, such value of stress is subtracted away from one of the two
local minima in Eq.~(\ref{eq:landscape2}). This leads to the slightly asymmetric yellow open circles fit by the parabolic dashed red curve, and whose coefficient is $F_1$.
The parameters of this model were then fed into our Monte Carlo solver to produce the critical temperatures in Fig.~\ref{fig:figure4}(c,f) later on.

\section{Discussion of group symmetries}\label{sec:appendixB}

Rodin and coworkers \cite{rodin_prb_2016_sns} identified the symmetry group of the ground state u.c.~[Fig.~\ref{fig:figure1}(a)] as Pnm$2_1$. The group has four generators which look as follows with the choice of principal axes in Fig.~\ref{fig:figure1}: $(x,y,z)$, $(x,-y,z)$, $(x+1/2,y+1/2,-z)$, and $(x+1/2,-y+1/2,-z)$. Since these materials have two atomic species, two labels for these fractional coordinates are needed. In order to be extremely explicit, we label them $x_{Sn}$, $y_{Sn}$, $z_{Sn}$, and $x_{Se}$, $y_{Se}$, $z_{Se}$, say for atoms $\mathbf{r}_{2,0}$ and $\mathbf{r}_{3,0}$, respectively. This way, $\mathbf{r}_{2,0}=\mathbf{a}_{1,0}x_{Sn}+\mathbf{a}_{2,0}y_{Sn}+\mathbf{a}_{3,0}z_{Sn}$ and $\mathbf{r}_{2,0}=\mathbf{a}_{1,0}x_{Se}+\mathbf{a}_{2,0}y_{Se}+\mathbf{a}_{3,0}z_{Se}$, with $\mathbf{a}_{3,0}$ sufficiently large to avoid interaction among periodic images. Being more specific, one writes $x_{Sn}=\delta_{x,0}/a_{1,0}$, $y_{Sn}=0$, $x_{Se}=0$, $y_{Se}=0$, and note that $z_{Sn}\ne -z_{Se}$ in principle. Atoms $\mathbf{r}_{1,0}$ and $\mathbf{r}_{4,0}$ can be obtained by applying the last two group generators (either a glide, or a two-fold rotation about the $\mathbf{a}_{1,0}$ axis), and the second group generator establishes a mirror symmetry about the $y-$axis.

Ref.~\onlinecite{fei_prl_2016} indicates that their paraelectric monolayer has a Cmcm symmetry. Such group is orthorhombic, which implies that the unit cell is rectangular. In fact, a Cmcm symmetry actually belongs to the bulk (see, e.g., \cite{SnSebulk3}) and not to monolayers: paraelectric monolayers with a rectangular unit cell have a Pnmm symmetry.
The demonstration goes as follows. Paraelectricity is obtained by setting both $x_{Sn}$ and $x_{Se}$ to zero since the paraelectric moment, resulting on an enhancement of symmetry with respect to the Pnm$2_1$ group by the creation of an additional reflection symmetry. The additional symmetry gives rise to four additional generators by the now allowed reflection about the $x-$axis of the generators from group Pnm2$_1$. Indeed, the eight generators turn out to be: $(x,y,z)$, $(x,-y,z)$, $(x+1/2,y+1/2,-z)$, $(x+1/2,-y+1/2,-z)$, $(-x,y,z)$, $(-x,-y,z)$, $(-x+1/2,y+1/2,-z)$, and $(-x+1/2,-y+1/2,-z)$. These generators belong to space group Pnmm.

Finally, once a paraelectric tetragonal u.c.~(i.e., one in which $a_1=a_2$) is permitted, then exchange among $x$ and $y$ components (representing a four-fold symmetry) is permitted; this enhanced symmetry then doubles the number of generators from those of the Pnmm group ($(x,y,z)$, $(x,-y,z)$, $(x+1/2,y+1/2,-z)$, $(x+1/2,-y+1/2,-z)$, $(-x,y,z)$, $(-x,-y,z)$, $(-x+1/2,y+1/2,-z)$, $(-x+1/2,-y+1/2,-z)$, $(y,x,z)$, $(y,-x,z)$, $(y+1/2,x+1/2,-z)$, $(y+1/2,-x+1/2,-z)$, $(-y,x,z)$, $(-y,-x,z)$, $(-y+1/2,x+1/2,-z)$, and $(-y+1/2,-x+1/2,-z)$), leading to Space Group P4/nmm.


\begin{thebibliography}{26}%
\makeatletter
\providecommand \@ifxundefined [1]{%
 \@ifx{#1\undefined}
}%
\providecommand \@ifnum [1]{%
 \ifnum #1\expandafter \@firstoftwo
 \else \expandafter \@secondoftwo
 \fi
}%
\providecommand \@ifx [1]{%
 \ifx #1\expandafter \@firstoftwo
 \else \expandafter \@secondoftwo
 \fi
}%
\providecommand \natexlab [1]{#1}%
\providecommand \enquote  [1]{``#1''}%
\providecommand \bibnamefont  [1]{#1}%
\providecommand \bibfnamefont [1]{#1}%
\providecommand \citenamefont [1]{#1}%
\providecommand \href@noop [0]{\@secondoftwo}%
\providecommand \href [0]{\begingroup \@sanitize@url \@href}%
\providecommand \@href[1]{\@@startlink{#1}\@@href}%
\providecommand \@@href[1]{\endgroup#1\@@endlink}%
\providecommand \@sanitize@url [0]{\catcode `\\12\catcode `\$12\catcode
  `\&12\catcode `\#12\catcode `\^12\catcode `\_12\catcode `\%12\relax}%
\providecommand \@@startlink[1]{}%
\providecommand \@@endlink[0]{}%
\providecommand \url  [0]{\begingroup\@sanitize@url \@url }%
\providecommand \@url [1]{\endgroup\@href {#1}{\urlprefix }}%
\providecommand \urlprefix  [0]{URL }%
\providecommand \Eprint [0]{\href }%
\providecommand \doibase [0]{https://doi.org/}%
\providecommand \selectlanguage [0]{\@gobble}%
\providecommand \bibinfo  [0]{\@secondoftwo}%
\providecommand \bibfield  [0]{\@secondoftwo}%
\providecommand \translation [1]{[#1]}%
\providecommand \BibitemOpen [0]{}%
\providecommand \bibitemStop [0]{}%
\providecommand \bibitemNoStop [0]{.\EOS\space}%
\providecommand \EOS [0]{\spacefactor3000\relax}%
\providecommand \BibitemShut  [1]{\csname bibitem#1\endcsname}%
\let\auto@bib@innerbib\@empty
\bibitem [{\citenamefont {Chang}\ \emph {et~al.}(2016)\citenamefont {Chang},
  \citenamefont {Liu}, \citenamefont {Lin}, \citenamefont {Wang}, \citenamefont
  {Zhao}, \citenamefont {Zhang}, \citenamefont {Jin}, \citenamefont {Zhong},
  \citenamefont {Hu}, \citenamefont {Duan}, \citenamefont {Zhang},
  \citenamefont {Fu}, \citenamefont {Xue}, \citenamefont {Chen},\ and\
  \citenamefont {Ji}}]{Kai}%
  \BibitemOpen
  \bibfield  {author} {\bibinfo {author} {\bibfnamefont {K.}~\bibnamefont
  {Chang}}, \bibinfo {author} {\bibfnamefont {J.}~\bibnamefont {Liu}}, \bibinfo
  {author} {\bibfnamefont {H.}~\bibnamefont {Lin}}, \bibinfo {author}
  {\bibfnamefont {N.}~\bibnamefont {Wang}}, \bibinfo {author} {\bibfnamefont
  {K.}~\bibnamefont {Zhao}}, \bibinfo {author} {\bibfnamefont {A.}~\bibnamefont
  {Zhang}}, \bibinfo {author} {\bibfnamefont {F.}~\bibnamefont {Jin}}, \bibinfo
  {author} {\bibfnamefont {Y.}~\bibnamefont {Zhong}}, \bibinfo {author}
  {\bibfnamefont {X.}~\bibnamefont {Hu}}, \bibinfo {author} {\bibfnamefont
  {W.}~\bibnamefont {Duan}}, \bibinfo {author} {\bibfnamefont {Q.}~\bibnamefont
  {Zhang}}, \bibinfo {author} {\bibfnamefont {L.}~\bibnamefont {Fu}}, \bibinfo
  {author} {\bibfnamefont {Q.-K.}\ \bibnamefont {Xue}}, \bibinfo {author}
  {\bibfnamefont {X.}~\bibnamefont {Chen}},\ and\ \bibinfo {author}
  {\bibfnamefont {S.-H.}\ \bibnamefont {Ji}},\ }\bibfield  {title} {\bibinfo
  {title} {\emph{{Discovery of Robust In-Plane Ferroelectricity in Atomic-Thick
  SnTe}}},\ }\href {https://doi.org/10.1126/science.aad8609} {\bibfield
  {journal} {\bibinfo  {journal} {Science}\ }\textbf {\bibinfo {volume}
  {353}},\ \bibinfo {pages} {274} (\bibinfo {year} {2016})}\BibitemShut
  {NoStop}%
\bibitem [{\citenamefont {Chang}\ \emph {et~al.}(2019)\citenamefont {Chang},
  \citenamefont {Miller}, \citenamefont {Yang}, \citenamefont {Lin},
  \citenamefont {Parkin}, \citenamefont {Barraza-Lopez}, \citenamefont {Xue},
  \citenamefont {Chen},\ and\ \citenamefont {Ji}}]{KaiPRL}%
  \BibitemOpen
  \bibfield  {author} {\bibinfo {author} {\bibfnamefont {K.}~\bibnamefont
  {Chang}}, \bibinfo {author} {\bibfnamefont {B.~J.}\ \bibnamefont {Miller}},
  \bibinfo {author} {\bibfnamefont {H.}~\bibnamefont {Yang}}, \bibinfo {author}
  {\bibfnamefont {H.}~\bibnamefont {Lin}}, \bibinfo {author} {\bibfnamefont
  {S.~S.~P.}\ \bibnamefont {Parkin}}, \bibinfo {author} {\bibfnamefont
  {S.}~\bibnamefont {Barraza-Lopez}}, \bibinfo {author} {\bibfnamefont {Q.-K.}\
  \bibnamefont {Xue}}, \bibinfo {author} {\bibfnamefont {X.}~\bibnamefont
  {Chen}},\ and\ \bibinfo {author} {\bibfnamefont {S.-H.}\ \bibnamefont {Ji}},\
  }\bibfield  {title} {\bibinfo {title} {\emph{{Standing Waves Induced by
  Valley-Mismatched Domains in Ferroelectric SnTe Monolayers}}},\ }\href
  {https://doi.org/10.1103/PhysRevLett.122.206402} {\bibfield  {journal}
  {\bibinfo  {journal} {Phys. Rev. Lett.}\ }\textbf {\bibinfo {volume} {122}},\
  \bibinfo {pages} {206402} (\bibinfo {year} {2019})}\BibitemShut {NoStop}%
\bibitem [{\citenamefont {Fei}\ \emph {et~al.}(2015)\citenamefont {Fei},
  \citenamefont {Li}, \citenamefont {Li},\ and\ \citenamefont
  {Yang}}]{fei_apl_2015_ges_gese_sns_snse}%
  \BibitemOpen
  \bibfield  {author} {\bibinfo {author} {\bibfnamefont {R.}~\bibnamefont
  {Fei}}, \bibinfo {author} {\bibfnamefont {W.}~\bibnamefont {Li}}, \bibinfo
  {author} {\bibfnamefont {J.}~\bibnamefont {Li}},\ and\ \bibinfo {author}
  {\bibfnamefont {L.}~\bibnamefont {Yang}},\ }\bibfield  {title} {\bibinfo
  {title} {\emph{{Giant Piezoelectricity of Monolayer Group {IV}
  Monochalcogenides: SnSe, SnS, GeSe, and GeS}}},\ }\href
  {https://doi.org/10.1063/1.4934750} {\bibfield  {journal} {\bibinfo
  {journal} {Appl. Phys. Lett.}\ }\textbf {\bibinfo {volume} {107}},\ \bibinfo
  {pages} {173104} (\bibinfo {year} {2015})}\BibitemShut {NoStop}%
\bibitem [{\citenamefont {Wang}\ and\ \citenamefont
  {Qian}(2017{\natexlab{a}})}]{wang_nanolett_2017_gese}%
  \BibitemOpen
  \bibfield  {author} {\bibinfo {author} {\bibfnamefont {H.}~\bibnamefont
  {Wang}}\ and\ \bibinfo {author} {\bibfnamefont {X.}~\bibnamefont {Qian}},\
  }\bibfield  {title} {\bibinfo {title} {\emph{{Giant Optical Second Harmonic
  Generation in Two-Dimensional Multiferroics}}},\ }\href
  {https://doi.org/10.1021/acs.nanolett.7b02268} {\bibfield  {journal}
  {\bibinfo  {journal} {Nano Lett.}\ }\textbf {\bibinfo {volume} {17}},\
  \bibinfo {pages} {5027} (\bibinfo {year} {2017}{\natexlab{a}})}\BibitemShut
  {NoStop}%
\bibitem [{\citenamefont {Panday}\ and\ \citenamefont {Fregoso}(2017)}]{b4}%
  \BibitemOpen
  \bibfield  {author} {\bibinfo {author} {\bibfnamefont {S.~R.}\ \bibnamefont
  {Panday}}\ and\ \bibinfo {author} {\bibfnamefont {B.~M.}\ \bibnamefont
  {Fregoso}},\ }\bibfield  {title} {\bibinfo {title} {\emph{{Strong Second
  Harmonic Generation in Two-Dimensional Ferroelectric
  IV-Monochalcogenides}}},\ }\href
  {http://stacks.iop.org/0953-8984/29/i=43/a=43LT01} {\bibfield  {journal}
  {\bibinfo  {journal} {J. Phys.: Condens. Matter}\ }\textbf {\bibinfo {volume}
  {29}},\ \bibinfo {pages} {43LT01} (\bibinfo {year} {2017})}\BibitemShut
  {NoStop}%
\bibitem [{\citenamefont {Shen}\ \emph {et~al.}(2019)\citenamefont {Shen},
  \citenamefont {Liu}, \citenamefont {Chang},\ and\ \citenamefont {Fu}}]{kai3}%
  \BibitemOpen
  \bibfield  {author} {\bibinfo {author} {\bibfnamefont {H.}~\bibnamefont
  {Shen}}, \bibinfo {author} {\bibfnamefont {J.}~\bibnamefont {Liu}}, \bibinfo
  {author} {\bibfnamefont {K.}~\bibnamefont {Chang}},\ and\ \bibinfo {author}
  {\bibfnamefont {L.}~\bibnamefont {Fu}},\ }\bibfield  {title} {\bibinfo
  {title} {\emph{{In-Plane Ferroelectric Tunnel Junction}}},\ }\href
  {https://doi.org/10.1103/PhysRevApplied.11.024048} {\bibfield  {journal}
  {\bibinfo  {journal} {Phys. Rev. Applied}\ }\textbf {\bibinfo {volume}
  {11}},\ \bibinfo {pages} {024048} (\bibinfo {year} {2019})}\BibitemShut
  {NoStop}%
\bibitem [{\citenamefont {Poudel}\ \emph {et~al.}(2019)\citenamefont {Poudel},
  \citenamefont {Villanova},\ and\ \citenamefont {Barraza-Lopez}}]{shiva}%
  \BibitemOpen
  \bibfield  {author} {\bibinfo {author} {\bibfnamefont {S.~P.}\ \bibnamefont
  {Poudel}}, \bibinfo {author} {\bibfnamefont {J.~W.}\ \bibnamefont
  {Villanova}},\ and\ \bibinfo {author} {\bibfnamefont {S.}~\bibnamefont
  {Barraza-Lopez}},\ }\bibfield  {title} {\bibinfo {title} {{Group-IV
  monochalcogenide monolayers: Two-dimensional ferroelectrics with weak
  intralayer bonds and a phosphorenelike monolayer dissociation energy}},\
  }\href {https://doi.org/10.1103/PhysRevMaterials.3.124004} {\bibfield
  {journal} {\bibinfo  {journal} {Phys. Rev. Materials}\ }\textbf {\bibinfo
  {volume} {3}},\ \bibinfo {pages} {124004} (\bibinfo {year}
  {2019})}\BibitemShut {NoStop}%
\bibitem [{\citenamefont {Mehboudi}\ \emph
  {et~al.}(2016{\natexlab{a}})\citenamefont {Mehboudi}, \citenamefont {Dorio},
  \citenamefont {Zhu}, \citenamefont {van~der Zande}, \citenamefont
  {Churchill}, \citenamefont {Pacheco-Sanjuan}, \citenamefont {Harriss},
  \citenamefont {Kumar},\ and\ \citenamefont {Barraza-Lopez}}]{Mehboudi2016}%
  \BibitemOpen
  \bibfield  {author} {\bibinfo {author} {\bibfnamefont {M.}~\bibnamefont
  {Mehboudi}}, \bibinfo {author} {\bibfnamefont {A.~M.}\ \bibnamefont {Dorio}},
  \bibinfo {author} {\bibfnamefont {W.}~\bibnamefont {Zhu}}, \bibinfo {author}
  {\bibfnamefont {A.}~\bibnamefont {van~der Zande}}, \bibinfo {author}
  {\bibfnamefont {H.~O.~H.}\ \bibnamefont {Churchill}}, \bibinfo {author}
  {\bibfnamefont {A.~A.}\ \bibnamefont {Pacheco-Sanjuan}}, \bibinfo {author}
  {\bibfnamefont {E.~O.}\ \bibnamefont {Harriss}}, \bibinfo {author}
  {\bibfnamefont {P.}~\bibnamefont {Kumar}},\ and\ \bibinfo {author}
  {\bibfnamefont {S.}~\bibnamefont {Barraza-Lopez}},\ }\bibfield  {title}
  {\bibinfo {title} {\emph{{Two-Dimensional Disorder in Black Phosphorus and
  Monochalcogenide Monolayers}}},\ }\href
  {https://doi.org/10.1021/acs.nanolett.5b04613} {\bibfield  {journal}
  {\bibinfo  {journal} {Nano Lett.}\ }\textbf {\bibinfo {volume} {16}},\
  \bibinfo {pages} {1704} (\bibinfo {year} {2016}{\natexlab{a}})}\BibitemShut
  {NoStop}%
\bibitem [{\citenamefont {Rodin}\ \emph {et~al.}(2016)\citenamefont {Rodin},
  \citenamefont {Gomes}, \citenamefont {Carvalho},\ and\ \citenamefont
  {Castro~Neto}}]{rodin_prb_2016_sns}%
  \BibitemOpen
  \bibfield  {author} {\bibinfo {author} {\bibfnamefont {A.~S.}\ \bibnamefont
  {Rodin}}, \bibinfo {author} {\bibfnamefont {L.~C.}\ \bibnamefont {Gomes}},
  \bibinfo {author} {\bibfnamefont {A.}~\bibnamefont {Carvalho}},\ and\
  \bibinfo {author} {\bibfnamefont {A.~H.}\ \bibnamefont {Castro~Neto}},\
  }\bibfield  {title} {\bibinfo {title} {\emph{{Valley Physics in Tin ({II})
  Sulfide}}},\ }\href {https://doi.org/10.1103/PhysRevB.93.045431} {\bibfield
  {journal} {\bibinfo  {journal} {Phys. Rev. B}\ }\textbf {\bibinfo {volume}
  {93}},\ \bibinfo {pages} {045431} (\bibinfo {year} {2016})}\BibitemShut
  {NoStop}%
\bibitem [{\citenamefont {Wu}\ and\ \citenamefont {Zeng}(2016)}]{otherarticle}%
  \BibitemOpen
  \bibfield  {author} {\bibinfo {author} {\bibfnamefont {M.}~\bibnamefont
  {Wu}}\ and\ \bibinfo {author} {\bibfnamefont {X.~C.}\ \bibnamefont {Zeng}},\
  }\bibfield  {title} {\bibinfo {title} {Intrinsic ferroelasticity and/or
  multiferroicity in two-dimensional phosphorene and phosphorene analogues},\
  }\href {https://doi.org/10.1021/acs.nanolett.6b00726} {\bibfield  {journal}
  {\bibinfo  {journal} {Nano Lett.}\ }\textbf {\bibinfo {volume} {16}},\
  \bibinfo {pages} {3236} (\bibinfo {year} {2016})}\BibitemShut {NoStop}%
\bibitem [{\citenamefont {Wang}\ and\ \citenamefont
  {Qian}(2017{\natexlab{b}})}]{other3}%
  \BibitemOpen
  \bibfield  {author} {\bibinfo {author} {\bibfnamefont {H.}~\bibnamefont
  {Wang}}\ and\ \bibinfo {author} {\bibfnamefont {X.}~\bibnamefont {Qian}},\
  }\bibfield  {title} {\bibinfo {title} {\emph{{Two-Dimensional Multiferroics
  in Monolayer Group IV Monochalcogenides}}},\ }\href
  {https://doi.org/10.1088/2053-1583/4/1/015042} {\bibfield  {journal}
  {\bibinfo  {journal} {2D Mater.}\ }\textbf {\bibinfo {volume} {4}},\ \bibinfo
  {pages} {015042} (\bibinfo {year} {2017}{\natexlab{b}})}\BibitemShut
  {NoStop}%
\bibitem [{\citenamefont {Potts}(1952)}]{potts}%
  \BibitemOpen
  \bibfield  {author} {\bibinfo {author} {\bibfnamefont {R.~B.}\ \bibnamefont
  {Potts}},\ }\bibfield  {title} {\bibinfo {title} {\emph{{Some Generalized
  Order-disorder Transformations}}},\ }\href
  {https://doi.org/10.1017/S0305004100027419} {\bibfield  {journal} {\bibinfo
  {journal} {Math. Proc. Cambridge Philos. Soc.}\ }\textbf {\bibinfo {volume}
  {48}},\ \bibinfo {pages} {106} (\bibinfo {year} {1952})}\BibitemShut
  {NoStop}%
\bibitem [{\citenamefont {Fei}\ \emph {et~al.}(2016)\citenamefont {Fei},
  \citenamefont {Kang},\ and\ \citenamefont {Yang}}]{fei_prl_2016}%
  \BibitemOpen
  \bibfield  {author} {\bibinfo {author} {\bibfnamefont {R.}~\bibnamefont
  {Fei}}, \bibinfo {author} {\bibfnamefont {W.}~\bibnamefont {Kang}},\ and\
  \bibinfo {author} {\bibfnamefont {L.}~\bibnamefont {Yang}},\ }\bibfield
  {title} {\bibinfo {title} {\emph{{Ferroelectricity and Phase Transitions in
  Monolayer Group-{IV} Monochalcogenides}}},\ }\href
  {https://doi.org/10.1103/PhysRevLett.117.097601} {\bibfield  {journal}
  {\bibinfo  {journal} {Phys. Rev. Lett.}\ }\textbf {\bibinfo {volume} {117}},\
  \bibinfo {pages} {097601} (\bibinfo {year} {2016})}\BibitemShut {NoStop}%
\bibitem [{\citenamefont {Barraza-Lopez}\ \emph {et~al.}(2018)\citenamefont
  {Barraza-Lopez}, \citenamefont {Kaloni}, \citenamefont {Poudel},\ and\
  \citenamefont {Kumar}}]{other4}%
  \BibitemOpen
  \bibfield  {author} {\bibinfo {author} {\bibfnamefont {S.}~\bibnamefont
  {Barraza-Lopez}}, \bibinfo {author} {\bibfnamefont {T.~P.}\ \bibnamefont
  {Kaloni}}, \bibinfo {author} {\bibfnamefont {S.~P.}\ \bibnamefont {Poudel}},\
  and\ \bibinfo {author} {\bibfnamefont {P.}~\bibnamefont {Kumar}},\ }\bibfield
   {title} {\bibinfo {title} {\emph{{Tuning the Ferroelectric-to-Paraelectric
  Transition Temperature and Dipole Orientation of Group-IV Monochalcogenide
  Monolayers}}},\ }\href {https://doi.org/10.1103/PhysRevB.97.024110}
  {\bibfield  {journal} {\bibinfo  {journal} {Phys. Rev. B}\ }\textbf {\bibinfo
  {volume} {97}},\ \bibinfo {pages} {024110} (\bibinfo {year}
  {2018})}\BibitemShut {NoStop}%
\bibitem [{\citenamefont {Soler}\ \emph {et~al.}(2002)\citenamefont {Soler},
  \citenamefont {Artacho}, \citenamefont {Gale}, \citenamefont {Garc{\'\i}a},
  \citenamefont {Junquera}, \citenamefont {Ordej{\'o}n},\ and\ \citenamefont
  {S\'anchez-Portal}}]{siesta}%
  \BibitemOpen
  \bibfield  {author} {\bibinfo {author} {\bibfnamefont {J.~M.}\ \bibnamefont
  {Soler}}, \bibinfo {author} {\bibfnamefont {E.}~\bibnamefont {Artacho}},
  \bibinfo {author} {\bibfnamefont {J.~D.}\ \bibnamefont {Gale}}, \bibinfo
  {author} {\bibfnamefont {A.}~\bibnamefont {Garc{\'\i}a}}, \bibinfo {author}
  {\bibfnamefont {J.}~\bibnamefont {Junquera}}, \bibinfo {author}
  {\bibfnamefont {P.}~\bibnamefont {Ordej{\'o}n}},\ and\ \bibinfo {author}
  {\bibfnamefont {D.}~\bibnamefont {S\'anchez-Portal}},\ }\bibfield  {title}
  {\bibinfo {title} {\emph{{The {SIESTA} Method for Ab Initio Order-{N}
  Materials Simulation}}},\ }\href
  {https://doi.org/10.1088/0953-8984/14/11/302} {\bibfield  {journal} {\bibinfo
   {journal} {J. Phys.: Condens. Matter}\ }\textbf {\bibinfo {volume} {14}},\
  \bibinfo {pages} {2745} (\bibinfo {year} {2002})}\BibitemShut {NoStop}%
\bibitem [{\citenamefont {Rom{\'a}n-P{\'e}rez}\ and\ \citenamefont
  {Soler}(2009)}]{soler}%
  \BibitemOpen
  \bibfield  {author} {\bibinfo {author} {\bibfnamefont {G.}~\bibnamefont
  {Rom{\'a}n-P{\'e}rez}}\ and\ \bibinfo {author} {\bibfnamefont {J.~M.}\
  \bibnamefont {Soler}},\ }\bibfield  {title} {\bibinfo {title}
  {\emph{{Efficient Implementation of a van der {Waals} Density Functional:
  Application to Double-Wall Carbon Nanotubes}}},\ }\href
  {https://doi.org/10.1103/PhysRevLett.103.096102} {\bibfield  {journal}
  {\bibinfo  {journal} {Phys. Rev. Lett.}\ }\textbf {\bibinfo {volume} {103}},\
  \bibinfo {pages} {096102} (\bibinfo {year} {2009})}\BibitemShut {NoStop}%
\bibitem [{\citenamefont {Berland}\ and\ \citenamefont {Hyldgaard}(2014)}]{BH}%
  \BibitemOpen
  \bibfield  {author} {\bibinfo {author} {\bibfnamefont {K.}~\bibnamefont
  {Berland}}\ and\ \bibinfo {author} {\bibfnamefont {P.}~\bibnamefont
  {Hyldgaard}},\ }\bibfield  {title} {\bibinfo {title} {Exchange functional
  that tests the robustness of the plasmon description of the van der waals
  density functional},\ }\href {https://doi.org/10.1103/PhysRevB.89.035412}
  {\bibfield  {journal} {\bibinfo  {journal} {Phys. Rev. B}\ }\textbf {\bibinfo
  {volume} {89}},\ \bibinfo {pages} {035412} (\bibinfo {year}
  {2014})}\BibitemShut {NoStop}%
\bibitem [{\citenamefont {Perdew}\ \emph {et~al.}(1996)\citenamefont {Perdew},
  \citenamefont {Burke},\ and\ \citenamefont {Ernzerhof}}]{PBE}%
  \BibitemOpen
  \bibfield  {author} {\bibinfo {author} {\bibfnamefont {J.~P.}\ \bibnamefont
  {Perdew}}, \bibinfo {author} {\bibfnamefont {K.}~\bibnamefont {Burke}},\ and\
  \bibinfo {author} {\bibfnamefont {M.}~\bibnamefont {Ernzerhof}},\ }\bibfield
  {title} {\bibinfo {title} {\emph{{Generalized Gradient Approximation Made
  Simple}}},\ }\href {https://doi.org/10.1103/PhysRevLett.77.3865} {\bibfield
  {journal} {\bibinfo  {journal} {Phys. Rev. Lett.}\ }\textbf {\bibinfo
  {volume} {77}},\ \bibinfo {pages} {3865} (\bibinfo {year}
  {1996})}\BibitemShut {NoStop}%
\bibitem [{\citenamefont {Mehboudi}\ \emph
  {et~al.}(2016{\natexlab{b}})\citenamefont {Mehboudi}, \citenamefont
  {Fregoso}, \citenamefont {Yang}, \citenamefont {Zhu}, \citenamefont {van~der
  Zande}, \citenamefont {Ferrer}, \citenamefont {Bellaiche}, \citenamefont
  {Kumar},\ and\ \citenamefont {Barraza-Lopez}}]{other2}%
  \BibitemOpen
  \bibfield  {author} {\bibinfo {author} {\bibfnamefont {M.}~\bibnamefont
  {Mehboudi}}, \bibinfo {author} {\bibfnamefont {B.~M.}\ \bibnamefont
  {Fregoso}}, \bibinfo {author} {\bibfnamefont {Y.}~\bibnamefont {Yang}},
  \bibinfo {author} {\bibfnamefont {W.}~\bibnamefont {Zhu}}, \bibinfo {author}
  {\bibfnamefont {A.}~\bibnamefont {van~der Zande}}, \bibinfo {author}
  {\bibfnamefont {J.}~\bibnamefont {Ferrer}}, \bibinfo {author} {\bibfnamefont
  {L.}~\bibnamefont {Bellaiche}}, \bibinfo {author} {\bibfnamefont
  {P.}~\bibnamefont {Kumar}},\ and\ \bibinfo {author} {\bibfnamefont
  {S.}~\bibnamefont {Barraza-Lopez}},\ }\bibfield  {title} {\bibinfo {title}
  {\emph{{Structural Phase Transition and Material Properties of Few-Layer
  Monochalcogenides}}},\ }\href
  {https://doi.org/10.1103/PhysRevLett.117.246802} {\bibfield  {journal}
  {\bibinfo  {journal} {Phys. Rev. Lett.}\ }\textbf {\bibinfo {volume} {117}},\
  \bibinfo {pages} {246802} (\bibinfo {year} {2016}{\natexlab{b}})}\BibitemShut
  {NoStop}%
\bibitem [{\citenamefont {Landau}(1941)}]{Landau2}%
  \BibitemOpen
  \bibfield  {author} {\bibinfo {author} {\bibfnamefont {L.}~\bibnamefont
  {Landau}},\ }\bibfield  {title} {\bibinfo {title} {\emph{{Theory of the
  Superfluidity of Helium II}}},\ }\href
  {https://doi.org/10.1103/PhysRev.60.356} {\bibfield  {journal} {\bibinfo
  {journal} {Phys. Rev.}\ }\textbf {\bibinfo {volume} {60}},\ \bibinfo {pages}
  {356} (\bibinfo {year} {1941})}\BibitemShut {NoStop}%
\bibitem [{\citenamefont {Bishop}\ and\ \citenamefont {Reppy}(1978)}]{BR1}%
  \BibitemOpen
  \bibfield  {author} {\bibinfo {author} {\bibfnamefont {D.~J.}\ \bibnamefont
  {Bishop}}\ and\ \bibinfo {author} {\bibfnamefont {J.~D.}\ \bibnamefont
  {Reppy}},\ }\bibfield  {title} {\bibinfo {title} {\emph{{Study of the
  superfluid transition in two-dimensional $^4$He films}}},\ }\href
  {https://doi.org/10.1103/PhysRevLett.40.1727} {\bibfield  {journal} {\bibinfo
   {journal} {Phys. Rev. Lett.}\ }\textbf {\bibinfo {volume} {40}},\ \bibinfo
  {pages} {1727} (\bibinfo {year} {1978})}\BibitemShut {NoStop}%
\bibitem [{\citenamefont {Kosterlitz}\ and\ \citenamefont
  {Thouless}(1973)}]{KostThou1973}%
  \BibitemOpen
  \bibfield  {author} {\bibinfo {author} {\bibfnamefont {J.~M.}\ \bibnamefont
  {Kosterlitz}}\ and\ \bibinfo {author} {\bibfnamefont {D.~J.}\ \bibnamefont
  {Thouless}},\ }\bibfield  {title} {\bibinfo {title} {\emph{{Ordering,
  Metastability and Phase Transitions in Two-Dimensional Systems}}},\ }\href
  {https://doi.org/10.1088/0022-3719/6/7/010} {\bibfield  {journal} {\bibinfo
  {journal} {J. Phys. C: Solid State Phys.}\ }\textbf {\bibinfo {volume} {6}},\
  \bibinfo {pages} {1181} (\bibinfo {year} {1973})}\BibitemShut {NoStop}%
\bibitem [{\citenamefont {Kosterlitz}(2016)}]{KostRev2016}%
  \BibitemOpen
  \bibfield  {author} {\bibinfo {author} {\bibfnamefont {J.~M.}\ \bibnamefont
  {Kosterlitz}},\ }\bibfield  {title} {\bibinfo {title}
  {\emph{{Kosterlitz-Thouless Physics: A Review of Key Issues}}},\ }\href
  {https://doi.org/10.1088/0034-4885/79/2/026001} {\bibfield  {journal}
  {\bibinfo  {journal} {Rep. Prog. Phys.}\ }\textbf {\bibinfo {volume} {79}},\
  \bibinfo {pages} {026001} (\bibinfo {year} {2016})}\BibitemShut {NoStop}%
\bibitem [{\citenamefont {Nahas}\ \emph {et~al.}(2017)\citenamefont {Nahas},
  \citenamefont {Prokhorenko}, \citenamefont {Kornev},\ and\ \citenamefont
  {Bellaiche}}]{Nahas2017}%
  \BibitemOpen
  \bibfield  {author} {\bibinfo {author} {\bibfnamefont {Y.}~\bibnamefont
  {Nahas}}, \bibinfo {author} {\bibfnamefont {S.}~\bibnamefont {Prokhorenko}},
  \bibinfo {author} {\bibfnamefont {I.}~\bibnamefont {Kornev}},\ and\ \bibinfo
  {author} {\bibfnamefont {L.}~\bibnamefont {Bellaiche}},\ }\bibfield  {title}
  {\bibinfo {title} {\emph{{Emergent Berezinskii-Kosterlitz-Thouless Phase in
  Low-Dimensional Ferroelectrics}}},\ }\href
  {https://doi.org/10.1103/PhysRevLett.119.117601} {\bibfield  {journal}
  {\bibinfo  {journal} {Phys. Rev. Lett.}\ }\textbf {\bibinfo {volume} {119}},\
  \bibinfo {pages} {117601} (\bibinfo {year} {2017})}\BibitemShut {NoStop}%
\bibitem [{\citenamefont {Zhu}\ \emph {et~al.}(2020)\citenamefont {Zhu},
  \citenamefont {Lu},\ and\ \citenamefont {Wang}}]{JAP}%
  \BibitemOpen
  \bibfield  {author} {\bibinfo {author} {\bibfnamefont {L.}~\bibnamefont
  {Zhu}}, \bibinfo {author} {\bibfnamefont {Y.}~\bibnamefont {Lu}},\ and\
  \bibinfo {author} {\bibfnamefont {L.}~\bibnamefont {Wang}},\ }\bibfield
  {title} {\bibinfo {title} {\emph{Tuning ferroelectricity by charge doping in
  two-dimensional SnSe}},\ }\href {https://doi.org/10.1063/1.5123296}
  {\bibfield  {journal} {\bibinfo  {journal} {J. Appl. Phys.}\ }\textbf
  {\bibinfo {volume} {127}},\ \bibinfo {pages} {014101} (\bibinfo {year}
  {2020})}\BibitemShut {NoStop}%
\bibitem [{\citenamefont {Li}\ \emph {et~al.}(2015)\citenamefont {Li},
  \citenamefont {Hong}, \citenamefont {May}, \citenamefont {Bansal},
  \citenamefont {Chi}, \citenamefont {Hong}, \citenamefont {Ehlers},\ and\
  \citenamefont {Delaire}}]{SnSebulk3}%
  \BibitemOpen
  \bibfield  {author} {\bibinfo {author} {\bibfnamefont {C.~W.}\ \bibnamefont
  {Li}}, \bibinfo {author} {\bibfnamefont {J.}~\bibnamefont {Hong}}, \bibinfo
  {author} {\bibfnamefont {A.~F.}\ \bibnamefont {May}}, \bibinfo {author}
  {\bibfnamefont {D.}~\bibnamefont {Bansal}}, \bibinfo {author} {\bibfnamefont
  {S.}~\bibnamefont {Chi}}, \bibinfo {author} {\bibfnamefont {T.}~\bibnamefont
  {Hong}}, \bibinfo {author} {\bibfnamefont {G.}~\bibnamefont {Ehlers}},\ and\
  \bibinfo {author} {\bibfnamefont {O.}~\bibnamefont {Delaire}},\ }\bibfield
  {title} {\bibinfo {title} {\emph{{Orbitally Driven Giant Phonon Anharmonicity
  in SnSe}}},\ }\href {https://doi.org/https://doi.org/10.1038/nphys3492}
  {\bibfield  {journal} {\bibinfo  {journal} {Nat. Phys.}\ }\textbf {\bibinfo
  {volume} {11}},\ \bibinfo {pages} {1063} (\bibinfo {year}
  {2015})}\BibitemShut {NoStop}%
\end{thebibliography}
\end{document}